\documentclass[
aps,
prd,
preprint,
superscriptaddress,
nofootinbib,
longbibliography,
floatfix
]{revtex4-2}

\usepackage{amsmath}
\usepackage{amssymb}
\usepackage{bm}
\usepackage{mathtools}
\usepackage{graphicx}
\usepackage{xcolor}
\usepackage{microtype}
\usepackage[hidelinks]{hyperref}

\usepackage{amsthm,booktabs}

\hypersetup{
  pdftitle={
    Testing a Flavor-Singlet Radial-Hybrid Interpretation of the X(2370):
    Glueball Ambiguity and Discriminating Amplitude Measurements
  },
  pdfauthor={Jianlong Lu},
  pdfsubject={
    Phenomenological comparison of radial-hybrid, glueball,
    mixed-glueball, and radial-quarkonium interpretations of the X(2370)
  },
  pdfkeywords={
    X(2370),
    pseudoscalar glueball,
    hybrid meson,
    radial excitation,
    radiative J/psi decay,
    amplitude analysis,
 conditional model comparison
  },
  pdfdisplaydoctitle=true
}

\begin{document}

\title{Testing a Flavor-Singlet Radial-Hybrid Interpretation of the
\texorpdfstring{$X(2370)$}{X(2370)}: Glueball Ambiguity and Discriminating
Amplitude Measurements}

\author{Jianlong Lu}
\email{jianlong@nus.edu.sg}
\affiliation{
Department of Mathematics, National University of Singapore,
Singapore 119076
}

\date{\today}

\begin{abstract}
The nature of the pseudoscalar $X(2370)$ remains unresolved: glueball,
hybrid, conventional $q\bar q$, and mixed configurations are all viable at
the level of present information.  We test a flavor-singlet
$J^{PC}=0^{-+}$ radial-hybrid interpretation using a channel-dependent
one-node overlap surrogate at $m_X=2.359~\mathrm{GeV}$ and a separate
flavor-singlet radiative
production factor.  The resulting decay template naturally suppresses
$K^{*}\bar K$ and can be conditionally normalized to the measured width,
although its dominant scalar and tensor
channels remain model dependent.  Reproducing the inferred central
$J/\psi\to\gamma X$ rate requires an effective singlet-production amplitude
about $4.4$ times the fixed-$V(0)$ continuation of the available $N_f=2$
lattice anchor.  In a common-nuisance screening comparison, the
hybrid--mixed-glueball integrated-score difference is $-1.29$, $+0.04$, and
$+0.41$ for tight, baseline, and wide nuisance settings, respectively; these
values constitute a conditional-score tie rather than a composition measurement.
We identify coherent amplitudes in $f_0(980)\eta$, $a_0(980)\pi$, and
$K_0^{*}(1430)\bar K$ as particularly informative.  Under the stated
proxy forecast covariance, including theory uncertainty for both templates,
their joint separation is $\Delta\chi^2\simeq6.5$ from the shared glueball
template and $\simeq7.6$ from the radial-$q\bar q$ surrogate.  A
flavor-singlet radial
hybrid is therefore compatible with current data but remains
indistinguishable from a mixed-glueball interpretation; a coherent amplitude
analysis can test the competing decay patterns.
\end{abstract}

\maketitle

\clearpage

\section{Introduction}
\label{sec:introduction}

The non-Abelian structure of quantum chromodynamics (QCD) permits hadronic
excitations in which gluonic degrees of freedom play an essential dynamical
role. Two especially important possibilities are glueballs, whose leading
valence configuration is purely gluonic, and hybrid mesons, in which a
quark--antiquark pair is coupled to an excitation of the gluon field.
Establishing either type of state would provide direct information about
confinement and the realization of low-energy gluodynamics. The experimental
identification is nevertheless difficult. Glueballs and nonexotic hybrids share
quantum numbers with conventional quarkonia, physical eigenstates can contain
several Fock components, and masses and individual branching fractions are not
composition observables by themselves. These difficulties are particularly
acute in the pseudoscalar-isoscalar sector, where the axial anomaly, flavor
mixing, radial nodes, and gluonic operators can all produce large effects
\cite{KlemptZaitsev2007HadronReview,MeyerSwanson2015HybridReview}.

Lattice-QCD calculations consistently place the lightest pseudoscalar glueball
in the approximate mass region $2.3$--$3.0~\mathrm{GeV}$, with representative
quenched determinations near $2.4$--$2.6~\mathrm{GeV}$
\cite{MorningstarPeardon1999,ChenEtAl2006Glueball,
GregoryEtAl2012Glueball,AthenodorouTeper2020Glueball}.
Radiative $J/\psi$ decays provide a natural environment in which to search
for such a state because the short-distance $c\bar c$ annihilation produces
a photon together with a gluon-rich hadronic system. In this environment the
BESIII Collaboration first observed the $X(2370)$ in
$J/\psi\to\gamma\pi^{+}\pi^{-}\eta^{\prime}$
\cite{BESIII2011X2370}. The state was subsequently confirmed in
$J/\psi\to\gamma K\bar K\eta^{\prime}$
\cite{BESIII2020X2370KKetaPrime}. A partial-wave analysis of
$J/\psi\to\gamma K_{S}^{0}K_{S}^{0}\eta^{\prime}$ later established
$J^{PC}=0^{-+}$ and measured the cascade product branching fraction through
$f_{0}(980)\eta^{\prime}$ \cite{BESIII2024X2370JPC}. The coincidence between
the measured mass and the lattice pseudoscalar-glueball window consequently
made the $X(2370)$ one of the leading glueball candidates.

The experimental picture has recently become substantially richer. Analyses
of the ten-billion-$J/\psi$ BESIII sample observed the $X(2370)$ in
$K_{S}^{0}K_{S}^{0}\pi^{0}$ and $\pi^{0}\pi^{0}\eta$, and observed
$X(2370)\to a_{0}(980)^{0}\pi^{0}$ with a statistical significance exceeding
$9\sigma$. Combining the available channels gives
\begin{equation}
 M_{X}=2359^{+13}_{-14}~\mathrm{MeV},
 \qquad
 \Gamma_{X}=170^{+44}_{-29}~\mathrm{MeV}.
\end{equation}
The combination is reported in Ref.~\cite{BESIII2026X2370NeutralModes}.
A subsequent search found no significant
$X(2370)\to K^{*}(892)^{0}\bar K^{0}+\mathrm{c.c.}$ signal and obtained
\begin{equation}
 \mathcal{B}\!\left[J/\psi\to\gamma X(2370)\right]
 \mathcal{B}\!\left[
 X(2370)\to K^{*}(892)^{0}\bar K^{0}+\mathrm{c.c.}
 \to K_{S}^{0}K_{S}^{0}\pi^{0}
 \right]
 <2.7\times10^{-6}
\end{equation}
at $90\%$ confidence level
\cite{BESIII2026X2370Glueball}. Together with the inclusive
$K\bar K\pi$, $\pi\pi\eta$, $\pi\pi\eta^{\prime}$, and
$K\bar K\eta^{\prime}$ product branching fractions, this suppression was
interpreted as evidence that the $X(2370)$ is a flavor-singlet state and,
more strongly, that its dominant constituent is the lightest pseudoscalar
glueball.

Several observations make the glueball interpretation compelling. The mass
lies within the lattice-QCD window, the state is produced prominently in a
gluon-rich radiative decay, and its observed three-pseudoscalar modes resemble
those of the $\eta_{c}$. Chiral effective theories also predict sizable
pseudoscalar-glueball decays into three pseudoscalars and into
scalar--pseudoscalar configurations
\cite{EshraimEtAl2013GlueballDecay,Eshraim2023GlueballVectors}. At the same
time, none of these properties is unique to a glueball. A quenched lattice
calculation gives
$\mathcal{B}(J/\psi\to\gamma G_{0^{-+}})=2.31(90)\times10^{-4}$,
which is substantially below phenomenological extractions of the
$X(2370)$ production rate \cite{GuiEtAl2019PseudoscalarGlueball,
SunEtAl2022NatureX2370}. Mixing between a pseudoscalar glueball and the
$c\bar c$ component associated with the $\eta_{c}$ can enhance the
radiative transition even for a small mixing angle, thereby providing a viable
mixed-glueball explanation \cite{ChenEtAl2026GlueballEtaCMixing}. Moreover, the
flavor-singlet $U_{A}(1)$ anomaly can enhance the coupling between gluons and
light pseudoscalar singlets \cite{JiangEtAl2023Eta2Radiative}. Radiative
production therefore probes a combination of gluonic content, quark
annihilation, anomaly effects, and mixing; it is not a direct measurement of
the glueball probability.

A flavor-singlet $0^{-+}$ hybrid supplies a particularly strong non-glueball
competitor. Nonperturbative lattice spectroscopy identifies a nonexotic
$0^{-+}$ member of the lightest hybrid supermultiplet, generated by a
chromomagnetic gluonic excitation coupled to a $q\bar q$ pair
\cite{Dudek2011HybridSupermultiplet,DudekEtAl2013IsoscalarHybrids}. Flux-tube,
constituent-gluon, and QCD-sum-rule calculations likewise support light
pseudoscalar hybrids and predict characteristic strong-decay patterns
\cite{PageSwansonSzczepaniak1999HybridDecays,FarinaSwanson2024HybridDecays,TanEtAl2024HybridSumRules}.
Most calculations place the lowest $0^{-+}$ hybrid below the $X(2370)$;
hence a hybrid interpretation of this state naturally requires a radial or
higher excitation, or a pole substantially shifted by hadronic dressing.
This qualification is important, but it does not exclude the hypothesis:
radial hybrid wave functions contain nodes, and their decay amplitudes need
not resemble those of the lowest state.

The signed constituent-gluon amplitudes contain an additional mechanism that
is directly relevant to the BESIII $K^{*}K$ limit. When the nonstrange and
strange hybrid components are combined near the flavor-singlet direction,
their $K^{*}K$ amplitudes interfere destructively
\cite{FarinaSwanson2024HybridDecays}. Thus, suppression of $K^{*}K$ can occur in a
flavor-singlet hybrid without requiring a glueball-dominated wave function.
For a radial hybrid, a node in the channel-dependent spatial overlap supplies
a second suppression mechanism. The present null result therefore establishes
compatibility with a flavor-singlet assignment, but it does not distinguish a
flavor cancellation from radial-node suppression or from a glueball decay
selection rule. The distinction must instead be sought in correlated channels
with related breakup momenta and different flavor coefficients, notably
$f_{0}(980)\eta$, $a_{0}(980)\pi$, and $K_{0}^{*}(1430)K$.

Conventional radial $q\bar q$ assignments also require consideration.
Calculations in the $^{3}P_{0}$ decay model have reached model-dependent
conclusions for high radial $\eta$- or $\eta'$-like assignments: some width
and channel patterns disfavor a pure conventional state, whereas another
parameter choice permits a fourth-radial assignment
\cite{LiuEtAl2010RadialPseudoscalar,YuEtAl2011RadialEta,
ChenPing2011RadialEta}.  These studies also show that radial nodes can strongly
alter individual amplitudes.  A meaningful comparison must therefore give
conventional $q\bar q$, hybrid, and glueball hypotheses
comparable freedom for radial structure, flavor breaking, production
normalization, and unobserved decay strength.

This requirement exposes the principal limitation of the existing literature.
Glueball, hybrid, and conventional-meson studies typically employ different
observables, different decay operators, and different nuisance assumptions.
Meanwhile, the presently published BESIII information consists primarily of
inclusive three-body product branching fractions and a suppressive
$K^{*}K$ constraint. The observation of an $a_{0}(980)\pi$ component does
not by itself provide its fit fraction, relative phase, or covariance with
$K_{0}^{*}(1430)K$, $f_{0}\eta^{(\prime)}$, and nonresonant amplitudes.
Consequently, treating the inclusive rates as incoherent sums of
quasi-two-body widths would discard interference information and could create
artificial discriminatory power.

In this work we perform a common, uncertainty-aware comparison of four
hypotheses: a flavor-singlet radial hybrid, a pure pseudoscalar glueball, a
glueball--$\eta_{c}$ mixed state, and a conventional radial $q\bar q$
state. For the hybrid hypothesis we construct a channel-dependent one-node SHO
surrogate at $M_{X}=2.359~\mathrm{GeV}$, retaining daughter-dependent scales
and final-state orbital momenta.  Only the light--strange interference sign
within $K^*K$ is anchored to the published constituent-gluon result; other
cross-channel signs are explicit assumptions. We implement a covariance-ready likelihood for
the decisive quasi-two-body channels, while explicitly excluding the
unreleased fractions from the present-data fit. We separately quantify the
flavor-singlet/anomaly matrix element required by
$J/\psi\to\gamma X(2370)$, and compare all four hypotheses using the same
mass, total-width, production, $K^{*}K$, and inclusive
three-pseudoscalar information with identical nuisance widths.

The resulting comparison does not identify a unique composition.  Defining
\(\Delta\ln Z=\ln Z_{\mathrm{hybrid}}-
\ln Z_{\mathrm{mixed\ glueball}}\), the benchmark analysis using the
asymmetric inclusive-channel covariance with \(\rho_{\rm sys}=0.5\) gives
\(\Delta\ln Z=+0.035\).  The tight and wide common-nuisance scenarios give
\(-1.289\) and \(+0.415\), respectively, and the ordering reverses under an
approximately \(2.5\%\) common narrowing of the baseline non-mass nuisance
widths.  An exact leave-one-observable-block analysis shows that this
near-degeneracy is not driven by the mass or total width.  At baseline,
radiative production together with the visible \(K^{*}K\) constraint
contributes \(-0.559\) to \(\Delta\ln Z\), while the inclusive
three-pseudoscalar composition contributes \(+0.710\).  Their partial
cancellation produces the conditional-score tie.  The principal conclusion is
therefore one of identifiability: a flavor-singlet radial hybrid is compatible
with current data, but remains indistinguishable from a mixed-glueball
interpretation under the same generic nuisance-width prescription.

The same framework identifies a concrete route to resolving the ambiguity.
Our radial-hybrid template predicts appreciable
$f_{0}(980)\eta$ and $K_{0}^{*}(1430)K$ fractions, while retaining a
strongly suppressed $K^{*}K$ channel. In a forecast with $20\%$ marginal
precision, correlation coefficient $0.3$, and a benchmark $50\%$
theory-uncertainty floor, the joint
$\{f_{0}(980)\eta,a_{0}(980)\pi,K_{0}^{*}(1430)K\}$ measurement gives
proxy-template distances of $\Delta\chi^{2}=6.52$ between the hybrid and
glueball templates and $\Delta\chi^{2}=7.61$ between the hybrid and
radial-$q\bar q$ templates. These numbers are forecasts rather than
measurements. They demonstrate that the next decisive experimental result is
not another isolated upper limit, but a coherent amplitude analysis reporting
fit fractions or complex amplitudes together with their statistical and
systematic covariance.

The remainder of this paper is organized as follows. In
Sec.~\ref{sec:experimental_inputs} we define the experimental data vector and
distinguish measured quantities from forecast inputs. In
Sec.~\ref{sec:common-model-comparison} we formulate the radial-hybrid, pure-glueball,
mixed-glueball, and conventional $q\bar q$ hypotheses. The
channel-dependent radial-overlap calculation is developed in
Sec.~\ref{sec:radial_hybrid_framework}, and radiative production, including singlet and
anomaly effects, is discussed in Sec.~\ref{sec:radiative-production}.
Section~\ref{subsec:common-likelihood} introduces the common likelihood and
nuisance treatment. Results, prior-sensitivity tests, and leave-one-observable
diagnostics are presented in Sec.~\ref{subsec:current-model-comparison-results}. The coherent-amplitude
measurements required to distinguish the competing interpretations are
identified in Sec.~\ref{sec:experimental-discrimination}, followed by the discussion
and conclusions in Secs.~\ref{sec:discussion} and \ref{sec:conclusions}.

\section{Experimental inputs and public-data boundary}
\label{sec:experimental_inputs}

The purpose of this section is to define the information used in the
present-data comparison and, equally importantly, the information that is not
yet publicly available.  We distinguish four classes of quantities:
\begin{enumerate}
    \item direct experimental measurements;
    \item phenomenologically inferred auxiliary inputs;
    \item quantities obtained by combining measurements with additional
          production, isospin, or width assumptions;
    \item projected quasi-two-body measurements used only in forecasts.
\end{enumerate}
Only the first two classes enter the current common-model likelihood.  Derived
conversions are quoted as diagnostics, while forecast observables are kept
strictly separate.  This distinction is essential because the published
three-body product branching fractions are not coherent quasi-two-body decay
fractions.

\subsection{Quantum numbers, mass, and total width}
\label{subsec:spectrum_input}

A partial-wave analysis of
$J/\psi\to\gamma K^0_S K^0_S\eta^\prime$ established
$J^{PC}=0^{-+}$ for the $X(2370)$
\cite{BESIII2024X2370JPC}.  We therefore compare only pseudoscalar
hypotheses in what follows.  The subsequent combination of the
$K^0_S K^0_S\pi^0$, $\pi^0\pi^0\eta$, and
$K^0_S K^0_S\eta^\prime$ measurements gives
\cite{BESIII2026X2370NeutralModes}
\begin{equation}
    M_X = 2.359^{+0.013}_{-0.014}\ {\rm GeV},
    \qquad
    \Gamma_X = 0.170^{+0.044}_{-0.029}\ {\rm GeV}.
    \label{eq:measured_mass_width}
\end{equation}
These are the line-shape parameters used throughout the analysis.  We retain
their asymmetric uncertainties rather than replacing them by a second average
over older, channel-dependent determinations.  For an observable $x$ with
quoted lower and upper errors, the experimental likelihood is represented by
the normalized split-normal density
\begin{equation}
 p_{\rm SN}(x\mid\mu,\sigma_-,\sigma_+)
 =
 \frac{\sqrt{2/\pi}}{\sigma_-+\sigma_+}
 \exp\!\left[
    -\frac{(x-\mu)^2}{2\sigma_s^2}
 \right],
 \qquad
 \sigma_s =
 \begin{cases}
    \sigma_-, & x<\mu,\\
    \sigma_+, & x\geq\mu.
 \end{cases}
 \label{eq:split_normal}
\end{equation}
This treatment does not imply that the quoted Breit--Wigner parameters are an
$S$-matrix pole determination; they are used as the common experimental
line-shape constraints available for all hypotheses.

\subsection{Inclusive three-pseudoscalar product branching fractions}
\label{subsec:inclusive_ppp}

The BESIII compilation reports four product branching fractions
\cite{BESIII2026X2370Glueball},
\begin{equation}
    {\cal P}_i
    \equiv
    {\cal B}(J/\psi\to\gamma X)
    {\cal B}(X\to i),
    \label{eq:product_bf_definition}
\end{equation}
summarized in Table~\ref{tab:inclusive_inputs}.  The first uncertainty is
statistical and the second is systematic.

\begin{table}[t]
    \centering
    \caption{
       Inclusive product branching fractions used in the present-data
       likelihood.  The quoted charge/isospin factors are those applied in
       the experimental compilation.  The $K\bar K\eta^\prime$ result
       combines the neutral and charged modes.
    }
    \label{tab:inclusive_inputs}
    \begin{tabular}{lcc}
        \toprule
        Final state $i$
        & ${\cal P}_i\;(10^{-4})$
        & Experimental extrapolation
        \\
        \midrule
        $K\bar K\pi$
        & $3.25\pm0.25{}^{+0.73}_{-0.75}$
        & isospin factor $12$
        \\
        $\pi\pi\eta$
        & $3.20\pm0.10{}^{+0.90}_{-1.00}$
        & isospin factor $3$
        \\
        $\pi\pi\eta^\prime$
        & $1.94\pm0.04{}^{+0.33}_{-0.88}$
        & isospin factor $1.5$
        \\
        $K\bar K\eta^\prime$
        & $0.39\pm0.05\pm0.10$
        & combined factors $4$ and $2$
        \\
        \bottomrule
    \end{tabular}
\end{table}

Their central sum is
\begin{equation}
    \sum_i {\cal P}_i = 8.78\times10^{-4}.
\end{equation}
Because the same radiative-production factor multiplies all four channels, we
use their normalized composition,
\begin{equation}
    f_i^{\rm inc}
    =
    \frac{{\cal P}_i}{\sum_j{\cal P}_j},
    \qquad
    \sum_i f_i^{\rm inc}=1.
    \label{eq:inclusive_composition}
\end{equation}
At the published central values,
\begin{equation}
 \mathbf f_{\rm central}^{\rm inc}
 =
 (0.3702,\;0.3645,\;0.2210,\;0.0444),
 \label{eq:inclusive_central_composition}
\end{equation}
where the channel ordering is
$(K\bar K\pi,\pi\pi\eta,\pi\pi\eta^\prime,
K\bar K\eta^\prime)$.  These quantities describe only the relative
composition of the four measured inclusive modes.  They must not be
interpreted as an exhaustive set of $X(2370)$ branching fractions.

To avoid a singular covariance caused by the unit-sum constraint, the
likelihood is constructed in additive-log-ratio coordinates,
\begin{equation}
 \mathbf y =
 \left(
   \ln\frac{f_{K\bar K\pi}}{f_{K\bar K\eta^\prime}},
   \ln\frac{f_{\pi\pi\eta}}{f_{K\bar K\eta^\prime}},
   \ln\frac{f_{\pi\pi\eta^\prime}}{f_{K\bar K\eta^\prime}}
 \right).
 \label{eq:alr_definition}
\end{equation}

No cross-channel systematic covariance matrix accompanies the four published
measurements.  We therefore do not claim to reconstruct the experimental
covariance.  Instead, we propagate the asymmetric uncertainties under an
explicit sensitivity model.  For Monte Carlo replica $n$, the common and
channel-specific systematic variates are combined as
\begin{align}
 u_i^{(n)}
 &=
 \sqrt{\rho_{\rm sys}}\,z_c^{(n)}
 +
 \sqrt{1-\rho_{\rm sys}}\,z_i^{(n)},                                      \\
 {\cal P}_i^{(n)}
 &=
   \mu_i
   +\sigma_{i,{\rm stat}}z_{i,{\rm stat}}^{(n)}
   +\sigma_{i,{\rm sys}}\!\left(u_i^{(n)}\right)u_i^{(n)}
 \quad\text{conditioned on }{\cal P}_j^{(n)}>0\ \forall j,
 \label{eq:asymmetric_ppp_sampling}
\end{align}
where all $z$'s are independent standard-normal variables before the common
component is introduced.  Here $\rho_{\rm sys}$ is the correlation of the
latent standardized systematic variates $u_i$, not the correlation of the
final asymmetric systematic shifts or of the measured observables.  We use
\begin{equation}
 \sigma_{i,{\rm sys}}(u)=
 \begin{cases}
   \sigma_{i,{\rm sys}}^-, & u<0,\\
   \sigma_{i,{\rm sys}}^+, & u\geq0.
 \end{cases}
\end{equation}
We reject an entire replica if any channel is nonpositive; no logarithmic
regulator is used.  At $\rho_{\rm sys}=0.5$, $1.44\%$ of proposed replicas are
rejected.  This is an explicit positive truncation of the sensitivity model,
not a reconstruction of the experimental likelihood.

Using $2.5\times10^5$ replicas and the benchmark
$\rho_{\rm sys}=0.5$, we obtain
\begin{equation}
 \langle\mathbf y\rangle
 =
 (2.1323,\;2.0863,\;1.4491),
 \label{eq:benchmark_alr_mean}
\end{equation}
and
\begin{equation}
 {\bf C}_{y} =
 \begin{pmatrix}
  0.1072 & 0.0661 & 0.0643\\
  0.0661 & 0.1408 & 0.0735\\
  0.0643 & 0.0735 & 0.2646
 \end{pmatrix}.
 \label{eq:benchmark_alr_covariance}
\end{equation}
The corresponding median normalized fractions and central $68\%$ intervals
are
\begin{equation}
 \mathbf f_{\rho_{\rm sys}=0.5}^{\rm inc}
 =
 \left(
  0.3788^{+0.0616}_{-0.0540},\;
  0.3713^{+0.0571}_{-0.0636},\;
  0.2090^{+0.0401}_{-0.0561},\;
  0.0456^{+0.0120}_{-0.0105}
 \right).
 \label{eq:benchmark_fraction_intervals}
\end{equation}
We repeat the complete model comparison for
$\rho_{\rm sys}=0,\;0.5,$ and $0.9$, rather than interpreting any one of
these assumed correlations as experimental information.

Unless stated otherwise, the benchmark current-data comparison uses
\(\rho_{\rm sys}=0.5\), \(N_{\rm rep}=2.5\times10^5\), and random seed
\(s=2870\).  The resulting additive-log-ratio mean and covariance in
Eqs.~\eqref{eq:benchmark_alr_mean} and
\eqref{eq:benchmark_alr_covariance} are passed without modification to the
common four-hypothesis likelihood in
Sec.~\ref{sec:common-model-comparison}.  The covariance-stress test varies
only \(\rho_{\rm sys}\); model anchors, nuisance distributions, likelihood
blocks, quadrature settings, and the missing-data gate are otherwise held
fixed.

\subsection{The visible \texorpdfstring{$K^\ast(892)\bar K$}{K-star Kbar} constraint}
\label{subsec:kstar_input}

The direct observable in the $K^\ast(892)^0\bar K^0+\mathrm{c.c.}$ search is
the visible cascade product branching fraction
\begin{align}
 {\cal P}_{K^\ast K}^{\rm vis}
 &\equiv
 {\cal B}(J/\psi\to\gamma X)
 {\cal B}\!\left[
  X\to K^\ast(892)^0\bar K^0+\mathrm{c.c.}
  \to K^0_S K^0_S\pi^0
 \right]                                                         \notag\\
 &=
 (0.1\pm1.2_{\rm stat}\pm1.1_{\rm syst})\times10^{-6}.
 \label{eq:kstar_visible_measurement}
\end{align}
The fitted signal significance is only $0.1\sigma$, and BESIII reports
\cite{BESIII2026X2370Glueball}
\begin{equation}
 {\cal P}_{K^\ast K}^{\rm vis}
 <2.7\times10^{-6}
 \qquad (90\%\ {\rm C.L.}).
 \label{eq:kstar_visible_ul}
\end{equation}
The corresponding ratio to the inclusive $K\bar K\pi$ product branching
fraction is
\begin{equation}
 R_{K^\ast K}
 =
 0.003\pm0.040_{\rm stat}\pm0.026_{\rm syst},
 \qquad
 R_{K^\ast K}<0.081
 \quad(90\%\ {\rm C.L.}).
 \label{eq:kstar_ratio}
\end{equation}
Because ${\cal P}_{K^\ast K}^{\rm vis}$ and $R_{K^\ast K}$ are obtained
from overlapping data, only the former is included in the common likelihood;
the ratio is retained as a validation observable.  Statistical and systematic
uncertainties in Eq.~\eqref{eq:kstar_visible_measurement} are combined in
quadrature,
\begin{equation}
 \sigma_{K^\ast K}^{\rm vis}
 =
 \sqrt{1.2^2+1.1^2}\times10^{-6}
 =
 1.628\times10^{-6}.
 \label{eq:kstar_combined_sigma}
\end{equation}
The published upper limit is used as an independent check of this Gaussian
likelihood treatment.

Converting Eq.~\eqref{eq:kstar_visible_ul} into a total
$X\to K^\ast\bar K$ branching fraction is not a direct experimental
operation.  It additionally requires a lower bound on radiative production
and a conversion from the measured neutral cascade to the full isospin-summed
mode.  Using
\begin{equation}
 {\cal B}(J/\psi\to\gamma X)>10^{-3},
 \qquad
 f_{\rm neutral}=\frac{1}{6},
\end{equation}
gives
\begin{equation}
 {\cal B}(X\to K^\ast\bar K)
 <
 \frac{2.7\times10^{-6}}
 {(10^{-3})(1/6)}
 =
 1.62\times10^{-2}.
 \label{eq:derived_kstar_bf}
\end{equation}
Combining this derived limit with the central width in
Eq.~\eqref{eq:measured_mass_width} gives
\begin{equation}
 \Gamma(X\to K^\ast\bar K)
 <
 (1.62\times10^{-2})(170~{\rm MeV})
 =
 2.75~{\rm MeV},
 \label{eq:derived_kstar_width}
\end{equation}
with a range $2.28$--$3.47~{\rm MeV}$ when only the quoted uncertainty on
the total width is varied.  We consequently use the directly measured visible
product branching fraction in the likelihood.  The total-mode branching and
partial-width bounds are reported only as assumption-dependent diagnostics,
and the commonly quoted ``below $2~{\rm MeV}$'' statement is not treated as
an additional independent measurement.

\subsection{Radiative-production information}
\label{subsec:production_input}

The four inclusive measurements determine products of production and decay
branching fractions; they do not separately measure
\begin{equation}
 b_\gamma\equiv{\cal B}(J/\psi\to\gamma X).
\end{equation}
For the production-sensitive part of the comparison, we use the two solutions
obtained from the phenomenological analysis of the measured invariant-mass
spectra by Sun \textit{et al.}\
\cite{SunEtAl2022NatureX2370}.  They are represented by the equal-weight auxiliary
likelihood
\begin{align}
 {\cal L}_{\gamma}^{\rm aux}(b_\gamma)
 &=
 \frac{1}{2}\,
 {\cal N}\!\left(
   b_\gamma;\,
   2.87\times10^{-3},\,
   0.68\times10^{-3}
 \right)                                                   \notag\\
 &\quad+
 \frac{1}{2}\,
 {\cal N}\!\left(
   b_\gamma;\,
   3.95\times10^{-3},\,
   0.71\times10^{-3}
 \right).
 \label{eq:production_mixture}
\end{align}
Equation~\eqref{eq:production_mixture} is not a standalone BESIII branching
fraction measurement.  It is a model-dependent extraction from exclusive
three-pseudoscalar spectra.  All production-sensitive score ratios reported
below are therefore conditional on this auxiliary likelihood.

The BESIII compilation also gives the cascade limits
\begin{equation}
 {\cal B}(J/\psi\to\gamma X\to\gamma\gamma\omega)
 <0.04\times10^{-6},
 \qquad
 {\cal B}(J/\psi\to\gamma X\to\gamma\gamma\phi)
 <0.11\times10^{-6},
 \label{eq:radiative_vector_limits}
\end{equation}
at $90\%$ confidence level
\cite{BESIII2026X2370Glueball}.  These limits are not included in the current
common likelihood because doing so requires a controlled electromagnetic-decay
calculation for every competing hypothesis.  They remain useful validation
observables for future extensions.

\subsection{Status of coherent quasi-two-body information}
\label{subsec:q2b_data_boundary}

The available analyses contain clear information about intermediate scalar
channels but not the covariance-ready observables required by our proposed
discrimination test.  In particular, the decay
\begin{equation}
 X(2370)\to a_0(980)^0\pi^0,
 \qquad
 a_0(980)^0\to\pi^0\eta,
\end{equation}
has been observed with a significance exceeding $9\sigma$
\cite{BESIII2026X2370NeutralModes}.  However, no coherent
$a_0(980)\pi$ fraction, complex coupling, or covariance with
$K_0^\ast(1430)\bar K$ has been released.

Similarly, the earlier partial-wave analysis measured
\cite{BESIII2024X2370JPC}
\begin{align}
 &{\cal B}(J/\psi\to\gamma X)\,
  {\cal B}(X\to f_0(980)\eta^\prime)\,
  {\cal B}(f_0(980)\to K^0_S K^0_S)
 \notag\\
 &\hspace{2.5cm}
 =
 \left(
  1.31\pm0.22_{\rm stat}
  {}^{+2.85}_{-0.84}{}_{\rm syst}
 \right)\times10^{-5}.
 \label{eq:f0_etaprime_cascade}
\end{align}
This is a related $f_0(980)\eta^\prime$ cascade product branching fraction,
not a measurement of the target $f_0(980)\eta$ fraction.

Fit fractions from a coherent amplitude analysis need not sum to unity because
interference terms are present.  Independent fractions with separately quoted
errors would therefore be insufficient for our purpose.  The required release
is either a set of complex amplitudes or a set of fit fractions accompanied by
their statistical and systematic covariance, efficiency information, and the
adopted interference convention.  Our preferred initial channel set is
\begin{equation}
 \left\{
  f_0(980)\eta\ \text{or}\ a_0(980)\pi,\;
  K_0^\ast(1430)\bar K
 \right\}.
 \label{eq:requested_q2b_set}
\end{equation}
The corresponding present-data entries are consequently set to
$\texttt{null}$.  No synthetic or forecast quasi-two-body fraction is used
in a current-data integrated score.

\subsection{Adopted data vector and covariance stress test}
\label{subsec:adopted_data_vector}

The information entering the current comparison can now be written as
\begin{align}
 {\cal D}_{\rm direct}
 &=
 \left\{
   M_X,\,
   \Gamma_X,\,
   \mathbf y_{\rm inc},\,
   {\cal P}_{K^\ast K}^{\rm vis}
 \right\},                                                     \\
 {\cal D}_{\rm auxiliary}
 &=
 \left\{
   {\cal L}_{\gamma}^{\rm aux}
 \right\},                                                     \\
 {\cal D}_{\rm missing}
 &=
 \left\{
   \mathbf f_{\rm q2b},\,
   {\bf C}_{\rm q2b}
 \right\}.
 \label{eq:data_partition}
\end{align}
The current-data likelihood is based on
${\cal D}_{\rm direct}\cup{\cal D}_{\rm auxiliary}$.
The quantities in ${\cal D}_{\rm missing}$ enter only the explicitly labeled
forecast analysis.

Table~\ref{tab:covariance_sensitivity} shows the change in the
hybrid-minus-mixed-glueball log-score difference after regenerating the inclusive
composition covariance from the asymmetric errors for three assumed systematic
correlations.  All other inputs and nuisance prescriptions are held fixed.

\begin{table}[t]
    \centering
    \caption{
      Sensitivity of the common comparison to the unknown cross-channel
      systematic correlation.  Every entry is obtained with the same model
      anchors, likelihood blocks, nuisance distributions, and numerical
      integration settings used in the headline comparison; only the
      inclusive-PPP covariance assumption is varied.  Positive values favor
      the radial-hybrid template and negative values favor the mixed-glueball
      template.  None of the displayed values of \(\rho_{\rm sys}\) is an
      experimentally measured correlation.
    }
    \label{tab:covariance_sensitivity}
    \begin{tabular}{cccc}
        \toprule
        \(\rho_{\rm sys}\)
        & tight nuisances
        & baseline nuisances
        & wide nuisances
        \\
        \midrule
        \(0.0\) & \(-1.532\) & \(+0.001\) & \(+0.407\)\\
        \(0.5\) & \(-1.289\) & \(+0.035\) & \(+0.415\)\\
        \(0.9\) & \(-1.033\) & \(+0.066\) & \(+0.422\)\\
        \bottomrule
    \end{tabular}
\end{table}

For the baseline nuisance model,
\begin{equation}
 0.0010
 \leq
 \log Z_H-\log Z_{G{\rm -mix}}
 \leq
 0.0658,
 \label{eq:baseline_covariance_range}
\end{equation}
which is a conditional-score tie throughout the covariance scan; the largest
baseline difference is only \(0.066\) log-score units.  By contrast, the tight
and wide nuisance scenarios continue
to prefer opposite hypotheses.  The hybrid--mixed-glueball ambiguity is
therefore not generated by one particular assumed cross-channel correlation.
It is controlled more strongly by the common theory and production nuisance
freedom.

\begin{figure}[t]
    \centering
    \includegraphics[width=\linewidth]{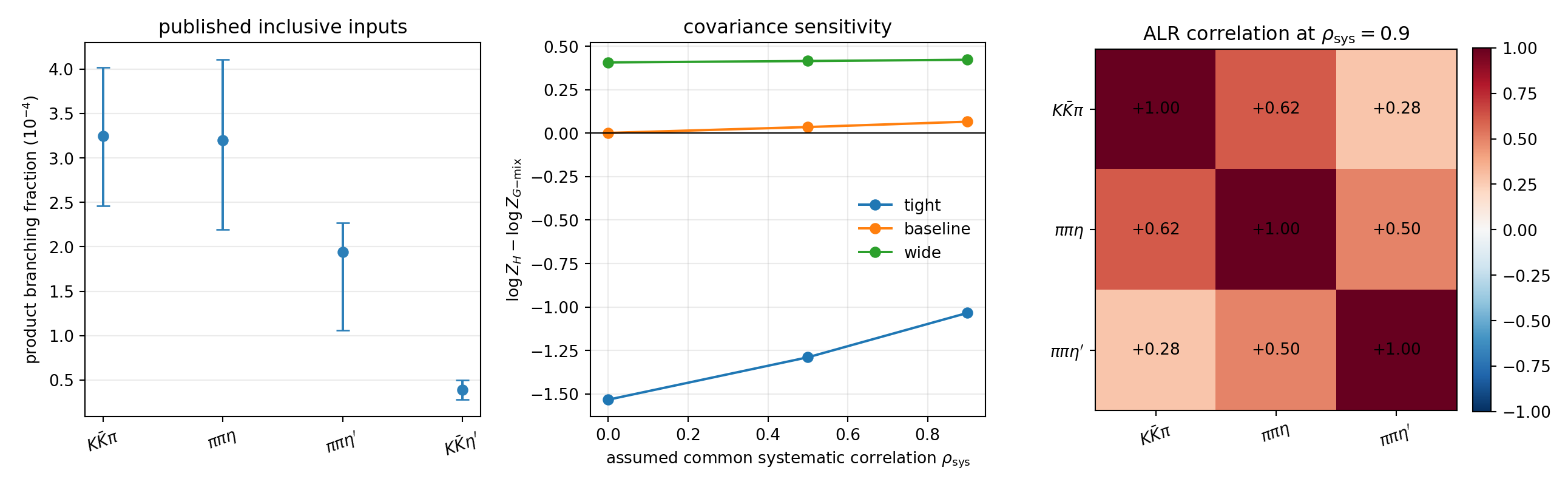}
    \caption{
  Experimental-input audit.  Left: the four published inclusive product
  branching fractions, with statistical and systematic uncertainties combined
  only for display.  Center: covariance sensitivity obtained by rerunning the
  same common four-model comparison used for the headline results while
  varying only the assumed cross-channel systematic correlation.
  Right: the induced additive-log-ratio correlation matrix for
  \(\rho_{\rm sys}=0.9\).  The correlation scan is a sensitivity test and
  not a reconstruction of an unpublished BESIII covariance matrix.
}
    \label{fig:experimental-inputs}
\end{figure}

This data boundary fixes the interpretation of all subsequent results.  The
present analysis tests whether a flavor-singlet radial-hybrid model is
compatible with the released measurements and whether it is distinguishable
from competing models under common nuisance freedom.  It does not use
unpublished amplitude information and does not convert the absence of such
information into evidence for any particular internal composition.

\section{Flavor-singlet radial-hybrid framework}
\label{sec:radial_hybrid_framework}

We now construct the flavor-singlet radial-hybrid hypothesis used in the
common comparison.  Lattice spectroscopy identifies a nonexotic
$0^{-+}$ state as a member of the lightest hybrid supermultiplet
\cite{Dudek2011HybridSupermultiplet,DudekEtAl2013IsoscalarHybrids}.
Consequently, the quantum numbers of the $X(2370)$ do not by themselves
distinguish a hybrid from a conventional pseudoscalar or a pseudoscalar
glueball.

The constituent-gluon calculation of Farina and Swanson provides signed
strong-decay amplitudes for the lowest light $0^{-+}$ hybrid
\cite{FarinaSwanson2024HybridDecays}.  Those amplitudes correspond to a state
near $1.75$--$1.90~{\rm GeV}$, not to a solved radial excitation at
$2.359~{\rm GeV}$.  We therefore use them as amplitude anchors and construct
the first radial excitation through explicit channel-dependent overlap
integrals.  The resulting calculation is more informative than multiplying
all channels by a common form factor, but it remains a separable
effective one-node reduction.  It is not a new solution of the complete
Coulomb-gauge three-body Hamiltonian.

\subsection{Flavor basis and singlet angle}
\label{subsec:hybrid_flavor_basis}

We define the light- and strange-quark hybrid basis states by
\begin{equation}
 \lvert H_n\rangle
 =
 \frac{\lvert u\bar ug\rangle+\lvert d\bar dg\rangle}{\sqrt{2}},
 \qquad
 \lvert H_s\rangle
 =
 \lvert s\bar sg\rangle,
 \label{eq:hybrid_flavor_basis}
\end{equation}
and parameterize an isoscalar hybrid as
\begin{equation}
 \lvert H(\theta_H)\rangle
 =
 \cos\theta_H\,\lvert H_n\rangle
 +
 \sin\theta_H\,\lvert H_s\rangle.
 \label{eq:hybrid_flavor_mixing}
\end{equation}
In this convention, the flavor-singlet state is
\begin{equation}
 \lvert H_1\rangle
 =
 \sqrt{\frac{2}{3}}\,\lvert H_n\rangle
 +
 \frac{1}{\sqrt{3}}\,\lvert H_s\rangle,
 \qquad
 \theta_1
 =
 \tan^{-1}\!\left(\frac{1}{\sqrt{2}}\right)
 =
 35.264^\circ.
 \label{eq:hybrid_singlet_angle}
\end{equation}

The light--strange relative sign within $K^*\bar K$ is taken from the
Farina--Swanson convention.  Their distinct-channel widths do not determine
physical phases between different final states.  At their reference mass
$M_0=1.750~{\rm GeV}$, the signed amplitudes relevant to the present analysis
can be written, in units of $\sqrt{\rm MeV}$, as
\begin{align}
 {\cal A}^{(0)}_{f_0(500)\eta}
 &=
 \sqrt{33.3}\cos\theta_H,
 &
 {\cal A}^{(0)}_{f_0(980)\eta}
 &=
 \sqrt{42.8}\cos\theta_H,
 \notag\\
 {\cal A}^{(0)}_{a_0(1450)\pi}
 &=
 \sqrt{34.5}\cos\theta_H,
 &
 {\cal A}^{(0)}_{a_2(1320)\pi}
 &=
 \cos\theta_H,
 \notag\\
 {\cal A}^{(0)}_{K^\ast\bar K}
 &=
 \sqrt{0.8}\cos\theta_H
 -
 \sqrt{2.0}\sin\theta_H.
 \label{eq:farina_swanson_anchor_amplitudes}
\end{align}
The minus sign in the last expression produces a flavor cancellation at
\begin{equation}
 \theta_{K^\ast K}^{(0)}
 =
 \tan^{-1}\!\sqrt{\frac{0.8}{2.0}}
 =
 32.312^\circ.
 \label{eq:kstar_exact_flavor_zero}
\end{equation}
This zero is only
\begin{equation}
 \theta_1-\theta_{K^\ast K}^{(0)}
 =
 2.953^\circ
 \label{eq:kstar_singlet_zero_separation}
\end{equation}
below the SU(3)-singlet angle.  Thus, the suppression of $K^\ast\bar K$
is already present in the signed lowest-hybrid amplitude and is not generated
by tuning the radial wave function.

\subsection{First-radial hybrid wave function}
\label{subsec:radial_hybrid_wavefunction}

The constituent-gluon construction represents the low-lying hybrid through a
transverse-electric quasigluon coupled to the $q\bar q$ pair.  Guided by the
Gaussian reduction used in Ref.~\cite{FarinaSwanson2024HybridDecays}, we
represent each momentum-space radial function by a normalized simple harmonic
oscillator form,
\begin{equation}
 R_{n\ell}(k;\beta)
 =
 {\cal N}_{n\ell}(\beta)
 \left(\frac{k}{\beta}\right)^\ell
 L_n^{\ell+1/2}\!\left(\frac{k^2}{\beta^2}\right)
 \exp\!\left(-\frac{k^2}{2\beta^2}\right),
 \label{eq:sho_radial_wavefunction}
\end{equation}
where
\begin{equation}
 {\cal N}_{n\ell}(\beta)
 =
 \left[
  \frac{2n!}
       {\beta^3\Gamma(n+\ell+3/2)}
 \right]^{1/2},
 \qquad
 \int_0^\infty dk\,k^2
 \left|R_{n\ell}(k;\beta)\right|^2=1.
 \label{eq:sho_normalization}
\end{equation}

The cited hybrid state depends on two Jacobi coordinates: the $q\bar q$
orbital motion and the transverse-electric gluon momentum.  The lowest
supermultiplet has $\ell_{q\bar q}=0$, while the gluonic factor carries one
power of momentum.  Our one-coordinate reduction does not retain that full
structure.  We assign the effective SHO factor
\begin{equation}
 n_H=1,
 \qquad
 \ell_H=1.
\end{equation}
Because
\begin{equation}
 L_1^{3/2}(x)=-x+\frac{5}{2},
\end{equation}
the parent wave function has an explicit momentum-space node at
\begin{equation}
 q_{\rm node}
 =
 \sqrt{\frac{5}{2}}\,\beta_H.
 \label{eq:first_radial_node}
\end{equation}
For the central value $\beta_H=0.50~{\rm GeV}$,
\begin{equation}
 q_{\rm node}=0.791~{\rm GeV}.
 \label{eq:central_radial_node}
\end{equation}
We vary
\begin{equation}
 0.35~{\rm GeV}
 \leq
 \beta_H
 \leq
 0.65~{\rm GeV},
 \label{eq:beta_scan_range}
\end{equation}
corresponding to
\begin{equation}
 0.553~{\rm GeV}
 \leq
 q_{\rm node}
 \leq
 1.028~{\rm GeV}.
 \label{eq:node_scan_range}
\end{equation}
This interval overlaps the breakup momenta of all principal channels and
therefore allows physically different channels to sample opposite sides of
the same radial node.

Ground-state daughter mesons are represented by $n=0$ SHO functions.
We use $\beta_\pi=\beta_\eta=0.40~{\rm GeV}$,
$\beta_K=0.42~{\rm GeV}$, vector-meson scales near
$0.38~{\rm GeV}$, and scalar/tensor scales in the range
$0.32$--$0.36~{\rm GeV}$.  These values define a controlled sensitivity
model; they are not fitted to the $X(2370)$ data.

\subsection{Channel-dependent overlap integral}
\label{subsec:hybrid_overlap_integral}

For a two-body channel $H\to A B$, let $p_i$ denote the relativistic
breakup momentum,
\begin{equation}
 p_i(M)
 =
 \frac{
  \sqrt{[M^2-(m_A+m_B)^2][M^2-(m_A-m_B)^2]}
 }{2M}.
 \label{eq:two_body_breakup_momentum}
\end{equation}
The mass-weighted momentum-sharing coefficients are
\begin{equation}
 \alpha_A=\frac{m_B}{m_A+m_B},
 \qquad
 \alpha_B=\frac{m_A}{m_A+m_B}.
 \label{eq:momentum_sharing_coefficients}
\end{equation}
For an integration momentum $q$ and
$\mu=\widehat{\boldsymbol q}\cdot\widehat{\boldsymbol p}$, the daughter
momenta entering their wave functions are
\begin{align}
 k_A^2
 &=
 q^2+\alpha_A^2p_i^2
 +2\alpha_A q p_i\mu,
 \notag\\
 k_B^2
 &=
 q^2+\alpha_B^2p_i^2
 -2\alpha_B q p_i\mu.
 \label{eq:daughter_internal_momenta}
\end{align}

After reducing the spin and color factors into the published anchor
amplitudes, the signed spatial overlap used here is
\begin{align}
 I_i^{(n)}(M,\beta_H)
 &=
 \int_0^{q_{\rm max}}dq\,q^2
 \int_{-1}^{1}d\mu\,
 R_{n1}(q;\beta_H)
 R_{0\ell_A}(k_A;\beta_A)
 R_{0\ell_B}(k_B;\beta_B)
 \notag\\
 &\quad\times
 q\,
 \exp\!\left(-\frac{q^2}{2\Lambda_{\rm op}^2}\right)
 P_{L_i}(\mu),
 \label{eq:radial_hybrid_overlap}
\end{align}
where $L_i$ is the relative final-state orbital angular momentum.  We use
\begin{equation}
 \Lambda_{\rm op}=0.90~{\rm GeV},
 \qquad
 q_{\rm max}=4.0~{\rm GeV}.
 \label{eq:operator_parameters}
\end{equation}
The extra factor of $q$ represents the reduced momentum dependence of the
leading quasigluon dissociation operator
$-g\int d^3x\,\psi^\dagger\boldsymbol{\alpha}\cdot
\boldsymbol A\psi$ used in the constituent-gluon model
\cite{FarinaSwanson2024HybridDecays}.  Equation~\eqref{eq:radial_hybrid_overlap}
preserves channel-dependent daughter masses, internal orbital assignments,
relative partial waves, nodes, and signs.

The integration uses 72-point Gauss--Legendre quadrature in $q$ and
56-point quadrature in $\mu$.  Increasing these orders to $96$ and $72$
changes the displayed overlaps by at most
\begin{equation}
 6.1\times10^{-6}
 \label{eq:overlap_numerical_convergence}
\end{equation}
in relative terms.  Numerical integration error is therefore negligible
compared with the physical model uncertainty.

\subsection{Transport of the published amplitudes}
\label{subsec:anchored_radial_amplitudes}

For a channel directly tabulated at $M_0$, the radial-hybrid amplitude at
$M_X$ is defined by
\begin{equation}
 \widetilde{\cal A}_i(M_X,\beta_H,\theta_H)
 =
 {\cal A}^{(0)}_i(M_0,\theta_H)
 \left[
  \frac{\Phi_i(M_X)}{\Phi_i(M_0)}
 \right]^{1/2}
 \frac{
  I_i^{(1)}(M_X,\beta_H)
 }{
  I_i^{(0)}(M_0,\beta_H)
 },
 \label{eq:anchored_radial_amplitude}
\end{equation}
with the exact two-body kernel
\begin{equation}
 \Phi_i(M)
 =
 \frac{p_i(M)^{2L_i+1}}{M^2}.
 \label{eq:phase_space_kernel}
\end{equation}
The corresponding anchored template weight is
\begin{equation}
 \widetilde\Gamma_i
 =
 \left|\widetilde{\cal A}_i\right|^2.
 \label{eq:anchored_template_width}
\end{equation}
The tilde emphasizes that these quantities are amplitude anchors modified by
our radial-overlap model, not precision absolute partial-width predictions.

Several important $S+P$ and tensor-pseudoscalar channels are closed or absent
from the published $1.75~{\rm GeV}$ table.  Following the characteristic
hybrid-decay preference for orbitally excited daughter mesons
\cite{PageSwansonSzczepaniak1999HybridDecays}, we construct these channels as
SU(3)-related partners.  If $a$ denotes a published anchor and $t$ its
partner, we use
\begin{equation}
 \widetilde{\cal A}_t
 =
 c_t\,\widetilde{\cal A}_a
 \left[
  \frac{N_t\Phi_t(M_X)}
       {N_a\Phi_a(M_X)}
 \right]^{1/2}
 \frac{I_t^{(1)}(M_X,\beta_H)}
      {I_a^{(1)}(M_X,\beta_H)},
 \label{eq:su3_partner_amplitude}
\end{equation}
where $N_i$ is the charge multiplicity and $c_t$ is an explicitly stated
reduced coupling.  The partner assignments are
\begin{align}
 a_0(1450)\pi
 &\longrightarrow
 K_0^\ast(1430)\bar K,
 &
 c_{K_0^\ast K}&=1,
 \notag\\
 a_0(1450)\pi
 &\longrightarrow
 f_0(1500)\eta,
 &
 c_{f_0(1500)\eta}
 &=
 0.75\cos15^\circ,
 \notag\\
 a_0(1450)\pi
 &\longrightarrow
 a_0(980)\pi,
 &
 c_{a_0(980)\pi}
 &=
 0.35,
 \notag\\
 a_2(1320)\pi
 &\longrightarrow
 K_2^\ast(1430)\bar K,
 &
 c_{K_2^\ast K}&=1,
 \notag\\
 a_2(1320)\pi
 &\longrightarrow
 f_2(1270)\eta,
 &
 c_{f_2(1270)\eta}
 &=
 \cos15^\circ.
 \label{eq:su3_partner_assignments}
\end{align}
These SU(3) relations, particularly the light-scalar coefficient in
$a_0(980)\pi$, are among the largest theory uncertainties in the hybrid
template.

\subsection{Normalization and decay-template observables}
\label{subsec:hybrid_template_normalization}

The relative decay template is
\begin{equation}
 f_i^{H}
 =
 \frac{\widetilde\Gamma_i}
      {\sum_j\widetilde\Gamma_j},
 \qquad
 \sum_i f_i^{H}=1.
 \label{eq:hybrid_normalized_fractions}
\end{equation}
These normalized quantities, rather than the unscaled absolute widths, enter
the decay-composition part of the common comparison.

For illustration only, one may impose saturation of the measured total width,
\begin{equation}
 \Gamma_i^{\rm sat}
 =
 \lambda_\Gamma^2\widetilde\Gamma_i,
 \qquad
 \lambda_\Gamma^2
 =
 \frac{170~{\rm MeV}}
      {\sum_j\widetilde\Gamma_j}.
 \label{eq:width_saturation_normalization}
\end{equation}
At $\beta_H=0.50~{\rm GeV}$,
\begin{equation}
 \sum_j\widetilde\Gamma_j=42.342~{\rm MeV},
 \qquad
 \lambda_\Gamma=2.004.
 \label{eq:central_width_rescaling}
\end{equation}
This rescaling does not assert that the modeled channels exhaust the physical
width.  It provides a convenient translation from normalized fractions into
conditional MeV values.

Table~\ref{tab:central_radial_hybrid_template} gives the complete central
template at the singlet angle.  The assumed-sign column records the adopted
real surrogate convention; only the light--strange relative sign internal to
$K^*K$ is source-anchored.  Only relative signs between amplitudes contributing
to the same physical final state can become observable in a coherent analysis.

\begin{table*}[t]
 \centering
 \caption{
  Central flavor-singlet radial-hybrid template at
  $\beta_H=0.50~{\rm GeV}$.  The last column assumes that the displayed
  channels saturate the measured $170~{\rm MeV}$ width and is therefore
  conditional rather than a direct prediction.
 }
 \label{tab:central_radial_hybrid_template}
 \begin{tabular}{lcccc}
  \toprule
  Channel
  & $p_i\;({\rm GeV})$
  & assumed sign
  & $100 f_i^H$
  & $\Gamma_i^{\rm sat}\;({\rm MeV})$
  \\
  \midrule
  $K^\ast(892)\bar K$      & 0.939 & $-$ & 0.0144 & 0.024 \\
  $f_0(500)\eta$           & 1.062 & $+$ & 4.38   & 7.44  \\
  $f_0(980)\eta$           & 0.879 & $+$ & 9.25   & 15.73 \\
  $a_0(1450)\pi$           & 0.732 & $+$ & 11.07  & 18.82 \\
  $a_0(980)\pi$            & 0.970 & $+$ & 0.442  & 0.752 \\
  $K_0^\ast(1430)\bar K$   & 0.629 & $+$ & 24.79  & 42.14 \\
  $f_0(1500)\eta$          & 0.515 & $+$ & 3.91   & 6.65  \\
  $a_2(1320)\pi$           & 0.803 & $+$ & 43.49  & 73.94 \\
  $K_2^\ast(1430)\bar K$   & 0.624 & $+$ & 2.07   & 3.52  \\
  $f_2(1270)\eta$          & 0.712 & $+$ & 0.576  & 0.979 \\
  \bottomrule
 \end{tabular}
\end{table*}

Two surrogate amplitude ratios illustrate the adopted convention:
\begin{equation}
 \left.
 \frac{
  \widetilde{\cal A}_{K^\ast\bar K}
 }{
  \widetilde{\cal A}_{K_0^\ast\bar K}
 }
 \right|_{\beta_H=0.50\,{\rm GeV}}
 =
 -0.0241,
 \label{eq:kstar_k0star_amplitude_ratio}
\end{equation}
and
\begin{equation}
 \left.
 \frac{
  \widetilde{\cal A}_{a_0(980)\pi}
 }{
  \widetilde{\cal A}_{f_0(980)\eta}
 }
 \right|_{\beta_H=0.50\,{\rm GeV}}
 =
 +0.219.
 \label{eq:a0_f0_amplitude_ratio}
\end{equation}
The $K^*\bar K$ light--strange cancellation is anchored, but its sign relative
to the SU(3)-partner $K_0^*(1430)\bar K$ amplitude is imposed.  Likewise, the
relative sign assigned to $a_0(980)\pi$ and $f_0(980)\eta$ is a surrogate-model
assumption, not an observable extracted from the published width table.  A
physical phase comparison requires predicted and measured complex couplings
in one common convention.

\subsection{Radial-scale and surrogate-sign stability}
\label{subsec:radial_hybrid_stability}

We scan $\beta_H$ uniformly over Eq.~\eqref{eq:beta_scan_range}.  The
resulting percentiles are sensitivity summaries of that scan, not posterior
credible intervals.  Table~\ref{tab:hybrid_beta_stability} gives the channels
most directly relevant to the proposed experimental test.

\begin{table}[t]
 \centering
 \caption{
  Radial-scale sensitivity of selected normalized fractions.  Entries are the
  16th, 50th, and 84th percentiles of a uniform
  $0.35\leq\beta_H\leq0.65~{\rm GeV}$ scan.
 }
 \label{tab:hybrid_beta_stability}
 \begin{tabular}{lccc}
  \toprule
  Channel
  & $100 f_{16}^H$
  & $100 f_{50}^H$
  & $100 f_{84}^H$
  \\
  \midrule
  $K^\ast(892)\bar K$
  & 0.0118 & 0.0137 & 0.0188
  \\
  $f_0(980)\eta$
  & 6.98 & 9.25 & 16.47
  \\
  $a_0(980)\pi$
  & 0.230 & 0.442 & 0.501
  \\
  $K_0^\ast(1430)\bar K$
  & 19.69 & 24.01 & 37.84
  \\
  \bottomrule
 \end{tabular}
\end{table}

Within the adopted surrogate convention, the $K^\ast\bar K$ and
$K_0^\ast(1430)\bar K$ amplitudes retain their assigned opposite signs over
the entire scan, and the assigned $f_0(980)\eta$ sign is stable.  Only the
light--strange relative sign internal to the $K^\ast\bar K$ anchor is inherited
from the published calculation; the cross-channel signs are assumptions.
The $a_0(980)\pi$ overlap passes through zero near
\begin{equation}
 \beta_H\simeq0.358~{\rm GeV},
 \label{eq:a0_phase_flip}
\end{equation}
and has the central positive sign over $96.7\%$ of the scanned interval.
This illustrates why an experimental phase convention and covariance matrix
are necessary: close to a radial zero, a fit fraction alone cannot distinguish
a sign change from an ordinary suppression.

The $a_2(1320)\pi$ contribution is both large and strongly
$\beta_H$-dependent.  Its normalized fraction spans approximately
\begin{equation}
 0.157
 \lesssim
 f_{a_2\pi}^H
 \lesssim
 0.541
 \label{eq:a2_fraction_sensitivity}
\end{equation}
between the 16th and 84th scan percentiles, and its amplitude changes sign near
$\beta_H\simeq0.386~{\rm GeV}$.  Consequently, the normalization of the
scalar fractions to the full modeled width remains substantially more
uncertain than the suppression of $K^\ast\bar K$.

\subsection{Predicted visible \texorpdfstring{$K^\ast\bar K$}{K-star Kbar} scale}
\label{subsec:hybrid_kstar_prediction}

At the singlet angle and central radial scale, the predicted normalized
$K^\ast\bar K$ fraction is
\begin{equation}
 f_{K^\ast K}^{H}
 =
 1.44\times10^{-4}.
 \label{eq:central_hybrid_kstar_fraction}
\end{equation}
If the displayed channels saturate the total width, this corresponds to
\begin{equation}
 \Gamma_{K^\ast K}^{\rm sat}
 =
 0.0244~{\rm MeV}.
 \label{eq:central_hybrid_kstar_width}
\end{equation}
For the working radiative-production value
${\cal B}(J/\psi\to\gamma H)=3.4\times10^{-3}$ and the visible neutral
fraction $f_{\rm neutral}=1/6$, the predicted product branching fraction is
\begin{align}
 {\cal P}_{K^\ast K}^{H,{\rm vis}}
 &=
 {\cal B}(J/\psi\to\gamma H)\,
 f_{K^\ast K}^{H}\,
 f_{\rm neutral}
 \notag\\
 &=
 8.14\times10^{-8}.
 \label{eq:central_hybrid_kstar_visible}
\end{align}
This is well below the direct experimental upper limit in
Eq.~\eqref{eq:kstar_visible_ul}.  The conclusion that a flavor-singlet radial
hybrid is compatible with the $K^\ast\bar K$ null result is therefore stable
within this construction.

The small number in Eq.~\eqref{eq:central_hybrid_kstar_visible} should not,
however, be interpreted as a unique hybrid signature.  It results from the
combination of an inherited near-singlet flavor zero and a channel-dependent
radial overlap.  A conventional radial $q\bar q$ state can also develop a
node, while a mixed-glueball model can suppress $K^\ast\bar K$ through a
different coupling structure.  The proposed coherent scalar-channel
measurements are required to distinguish these mechanisms.

\begin{figure*}[t]
 \centering
 \includegraphics[width=\textwidth]{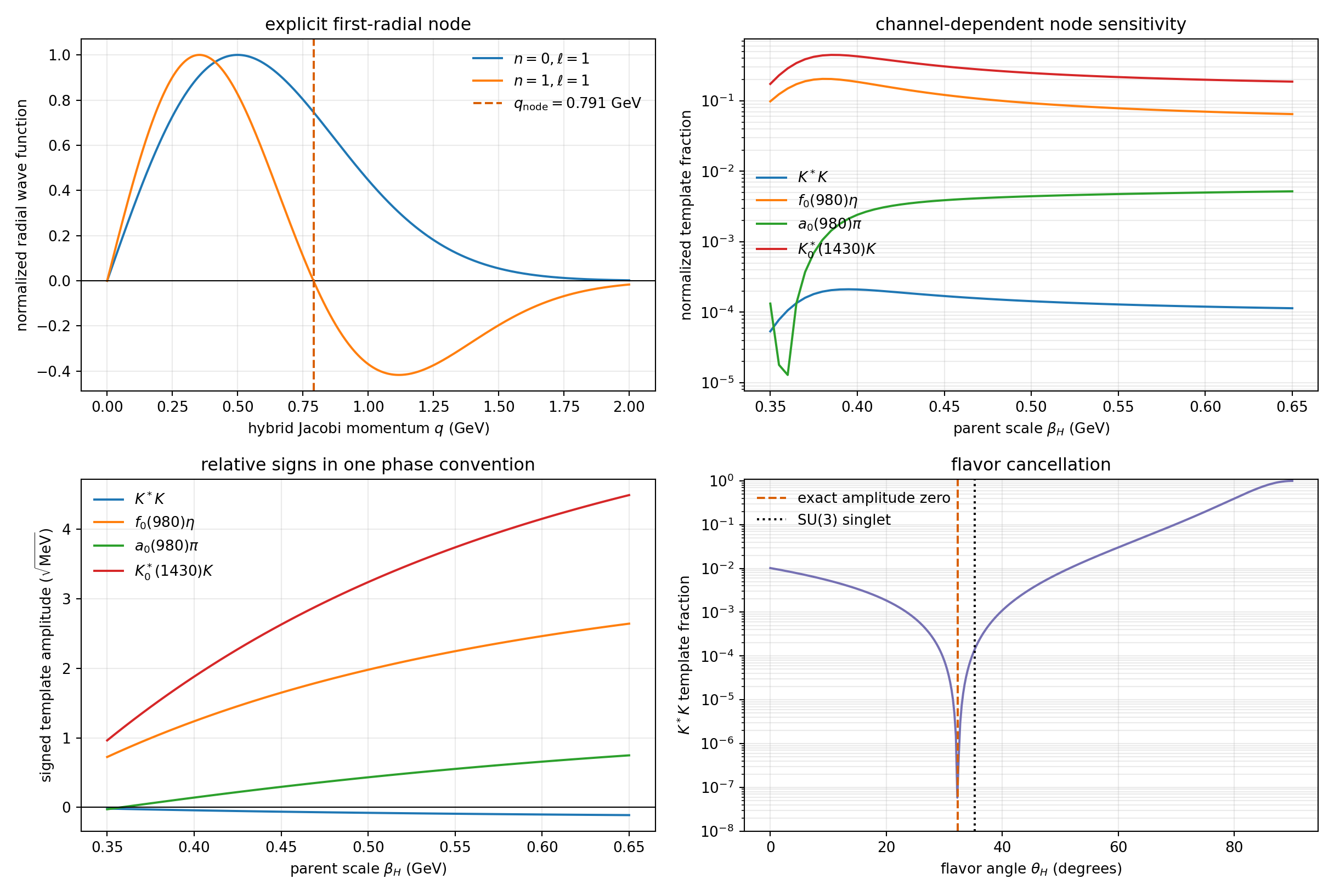}
 \caption{
  Anatomy of the radial-hybrid decay template.  Upper left: normalized
  $n=0$ and $n=1$, $\ell=1$ SHO radial functions at
  $\beta_H=0.50~{\rm GeV}$, including the explicit radial node.  Upper
  right: selected normalized fractions under variation of $\beta_H$.
  Lower left: signed amplitudes in the adopted real phase convention.
  Lower right: the $K^\ast\bar K$ flavor cancellation, showing the exact
  reference-amplitude zero and the SU(3)-singlet angle.
 }
 \label{fig:radial_hybrid_formalism}
\end{figure*}

\subsection{Domain of validity}
\label{subsec:radial_hybrid_validity}

The hybrid calculation used in this work has the following explicit
limitations:
\begin{enumerate}
 \item The radial wave function is an effective one-coordinate, one-node SHO
       surrogate, neither the full two-Jacobi-coordinate wave function nor a
       solution of the constituent-gluon Hamiltonian at $2.359~{\rm GeV}$.

 \item The absolute normalization is inherited from amplitudes calculated for
       the lowest hybrid at $M_0=1.750~{\rm GeV}$.  The optional normalization
       to $170~{\rm MeV}$ assumes saturation by the displayed channels.

 \item Channels that are absent or closed at the reference mass are generated
       through approximate SU(3)-partner relations.  Their reduced couplings,
       especially those involving the light scalar nonet, are model inputs.

 \item Nominal daughter masses are used in the overlap surrogate.  Broad
       scalar and tensor line shapes, coupled-channel dressing, three-body
       unitarity, and final-state rescattering are not solved at this stage.

 \item Only the light--strange sign inside the $K^*K$ anchor is inherited.
       Cross-channel and SU(3)-partner signs are assumptions.  Experimental
       comparison requires complex amplitudes and their covariance.

 \item The $\beta_H$ scan probes radial-size sensitivity but is not a complete
       theory covariance.  Daughter wave-function scales, scalar assignments,
       SU(3) breaking, and the decay operator require additional nuisance
       freedom in the common comparison.
\end{enumerate}
Accordingly, this framework supports a compatibility statement: a
flavor-singlet radial hybrid naturally accommodates the observed
$K^\ast\bar K$ suppression and gives conditional intensity targets.
It does not, by itself, establish that the $X(2370)$ is predominantly a
hybrid.

\section{Radiative \texorpdfstring{$J/\psi$}{J/psi} production and the
flavor-singlet anomaly}
\label{sec:radiative-production}

Radiative production in $J/\psi$ decay is frequently regarded as evidence
for gluonic structure.  Such an interpretation is plausible because the
annihilation of the initial $c\bar c$ pair produces a photon together with a
gluon-rich light-hadron system.  It is not, however, a composition projector.
A flavor-singlet $q\bar q g$ hybrid can couple through disconnected
light-quark diagrams and the axial anomaly, while a predominantly glueball
state can acquire a disproportionately large production amplitude from a small
$c\bar c$ admixture.  We therefore use radiative production as a quantitative
matrix-element constraint and not as a direct measurement of the hybrid,
glueball, or $q\bar q$ probability.

The presently quoted values of
$\mathcal B(J/\psi\to\gamma X)$ are not direct inclusive measurements.
They are obtained by dividing exclusive product branching fractions by
model-dependent estimates of the corresponding $X(2370)$ decay fractions.
The two solutions found in the three-pseudoscalar analysis are
\begin{align}
 {\cal B}_{\gamma}^{\rm I}
 &=
 (2.87\pm0.68)\times10^{-3},
 &
 {\cal B}_{\gamma}^{\rm II}
 &=
 (3.95\pm0.71)\times10^{-3},
 \label{eq:production-solutions}
\end{align}
where
${\cal B}_{\gamma}\equiv{\cal B}(J/\psi\to\gamma X)$
\cite{SunEtAl2022NatureX2370}.  These solutions represent alternative
treatments of the hadronic decay dynamics and must not be multiplied as
independent measurements.  We use
\begin{equation}
 {\cal B}_{\gamma}^{\rm work}=3.40\times10^{-3}
 \label{eq:production-working-point}
\end{equation}
only as a transparent working point for the following amplitude diagnostics.
The two solutions in Eq.~\eqref{eq:production-solutions} are retained
separately in the model-comparison likelihood.

\subsection{Transition convention and exact kinematics}
\label{subsec:radiative-convention}

For a pseudoscalar state $P$, we define the on-shell transition form factor by
\begin{equation}
 \left\langle P(k)\left|j_c^\mu(0)\right|
 J/\psi(p,\lambda)\right\rangle
 =
 {\cal F}_{P}(Q^2)\,
 \epsilon^{\mu\nu\rho\sigma}
 p_\nu k_\rho
 \epsilon_\sigma^{(\lambda)}(p),
 \qquad
 Q^2=-(p-k)^2 ,
 \label{eq:pseudoscalar-transition-definition}
\end{equation}
where $j_c^\mu=\bar c\gamma^\mu c$.  With the charm electric charge included
in the normalization, the real-photon width is
\begin{equation}
 \Gamma(J/\psi\to\gamma P)
 =
 \frac{4\alpha}{27}\,
 k_{\gamma}^{\,3}
 \left|{\cal F}_{P}(0)\right|^2,
 \qquad
 k_{\gamma}
 =
 \frac{M_{\psi}^{2}-M_{P}^{2}}{2M_{\psi}} .
 \label{eq:pseudoscalar-radiative-width}
\end{equation}
We take
\begin{equation}
 M_\psi=3.096900~{\rm GeV},
 \qquad
 \Gamma_\psi=92.6~{\rm keV},
 \qquad
 \alpha^{-1}=134 .
 \label{eq:radiative-numerical-inputs}
\end{equation}
The $J/\psi$ mass and width are taken from the Particle Data Group
\cite{NavasEtAl2024PDG}; $\alpha^{-1}=134$ follows the effective convention
of the lattice transition analyses cited below rather than the Thomson-limit
fine-structure constant.
The dimensionful form factor in
Eq.~\eqref{eq:pseudoscalar-radiative-width} is related to the commonly quoted
dimensionless lattice form factor by
\begin{equation}
 {\cal F}_{P}(0)
 =
 \frac{2\omega_P V_P(0)}{M_\psi+M_P},
 \label{eq:dimensionless-radiative-form-factor}
\end{equation}
where $\omega_P=1$ for the glueball-like annihilation convention and
$\omega_P=2$ for the connected $c\bar c$ transition convention
\cite{GuiEtAl2019PseudoscalarGlueball,
ChenEtAl2026GlueballEtaCMixing}.

At the central $X(2370)$ mass used in this work,
\begin{equation}
 M_X=2.359~{\rm GeV},
 \qquad
 k_{\gamma X}=0.649990~{\rm GeV}.
 \label{eq:x-photon-momentum}
\end{equation}
Inverting Eq.~\eqref{eq:pseudoscalar-radiative-width} gives the matrix element
required by any assumed production branching fraction,
\begin{equation}
 \left|{\cal F}_{X}^{\rm req}(0)\right|
 =
 \left[
 \frac{27\,\Gamma_\psi\,{\cal B}_\gamma}
      {4\alpha\,k_{\gamma X}^{3}}
 \right]^{1/2}.
 \label{eq:required-radiative-form-factor}
\end{equation}
The resulting values are displayed in
Table~\ref{tab:required-radiative-form-factor}.

\begin{table}[t]
 \centering
 \caption{
 Required $J/\psi\to\gamma X$ transition matrix element at
 $M_X=2.359~{\rm GeV}$.  The working point is a diagnostic midpoint,
 not a third measurement or a statistical average of solutions I and II.
 }
 \label{tab:required-radiative-form-factor}
 \begin{tabular}{lcc}
  \toprule
  Production input
  & ${\cal B}_\gamma$
  & $\left|{\cal F}_{X}^{\rm req}(0)\right|$
    $[\mathrm{GeV}^{-1}]$ \\
  \midrule
  Solution I
  & $(2.87\pm0.68)\times10^{-3}$
  & $0.02959$ \\
  Working point
  & $3.40\times10^{-3}$
  & $0.03220$ \\
  Solution II
  & $(3.95\pm0.71)\times10^{-3}$
  & $0.03471$ \\
  \bottomrule
 \end{tabular}
\end{table}

Thus, the phenomenological production range can be restated without assuming
a constituent interpretation:
\begin{equation}
 \left|{\cal F}_{X}(0)\right|
 \simeq (0.030\text{--}0.035)~{\rm GeV}^{-1}.
 \label{eq:required-form-factor-range}
\end{equation}
This matrix-element requirement is the central result of the present
radiative-production analysis.

\subsection{Flavor-singlet anomaly anchor}
\label{subsec:anomaly-anchor}

The divergence of the flavor-singlet axial current contains the topological
gluon operator,
\begin{equation}
 \partial_\mu J_{5}^{\mu,0}
 =
 2i\sum_{q=1}^{N_f}m_q\bar q\gamma_5q
 +
 \frac{N_f\alpha_s}{4\pi}
 G_{\mu\nu}^{a}\widetilde G^{a\mu\nu},
 \label{eq:singlet-axial-anomaly}
\end{equation}
up to the normalization adopted for $J_{5}^{\mu,0}$.  Consequently, a
flavor-singlet pseudoscalar can couple strongly to the gluonic system generated
in $c\bar c$ annihilation even when an analogous flavor-octet transition is
OZI suppressed.

The available first-principles anchor is the $N_f=2$ lattice calculation of
the light pseudoscalar singlet $\eta_{(2)}$.  It finds
\begin{align}
 m_{\eta_{(2)}} &= 0.718(8)~{\rm GeV},
 &
 \Gamma(J/\psi\to\gamma\eta_{(2)})
 &=0.385(45)~{\rm keV},
 \label{eq:eta2-lattice-anchor}
\end{align}
at $m_\pi\simeq350~{\rm MeV}$
\cite{JiangEtAl2023Eta2Radiative}.  Equation~\eqref{eq:pseudoscalar-radiative-width}
then gives
\begin{equation}
 \left|{\cal F}_{\eta_{(2)}}(0)\right|
 =0.01052~{\rm GeV}^{-1}.
 \label{eq:eta2-dimensional-form-factor}
\end{equation}

This result demonstrates anomaly-enhanced singlet production, but it is not a
calculation of a $2.359~{\rm GeV}$ radial hybrid.  In particular, a continuation
in the final-state mass requires an assumption about which form-factor
convention is held fixed.  We therefore display two deliberately simple
continuations:
\begin{subequations}
\label{eq:singlet-continuations}
\begin{align}
 {\rm fixed}\ {\cal F}: \qquad
 {\cal F}_{\eta_{(2)}}^{[{\cal F}]}(M_X)
 &=
 {\cal F}_{\eta_{(2)}}(m_{\eta_{(2)}}),
 \label{eq:fixed-dimensional-continuation}
 \\
 {\rm fixed}\ V: \qquad
 {\cal F}_{\eta_{(2)}}^{[V]}(M_X)
 &=
 {\cal F}_{\eta_{(2)}}(m_{\eta_{(2)}})
 \frac{M_\psi+m_{\eta_{(2)}}}{M_\psi+M_X}.
 \label{eq:fixed-dimensionless-continuation}
\end{align}
\end{subequations}
They give
\begin{align}
 {\cal B}_{\gamma}^{[{\cal F}]}
 &=3.63\times10^{-4},
 &
 {\cal B}_{\gamma}^{[V]}
 &=1.77\times10^{-4}.
 \label{eq:singlet-continuation-branching-fractions}
\end{align}
The difference between these numbers is convention dependence associated with
the mass continuation.  It must not be interpreted as a calculated
radial-hybrid theory uncertainty.

Relative to the working rate, the corresponding amplitude enhancements are
\begin{align}
 {\cal E}_{[{\cal F}]}
 &\equiv
 \frac{\left|{\cal F}_{X}^{\rm req}\right|}
      {\left|{\cal F}_{\eta_{(2)}}^{[{\cal F}]}\right|}
 =3.06,
 &
 {\cal E}_{[V]}
 &\equiv
 \frac{\left|{\cal F}_{X}^{\rm req}\right|}
      {\left|{\cal F}_{\eta_{(2)}}^{[V]}\right|}
 =4.38.
 \label{eq:singlet-required-enhancements}
\end{align}
Sampling the two production solutions with equal weight and propagating the
published $\eta_{(2)}$ width uncertainty gives, for the fixed-$V$
continuation,
\begin{equation}
 {\cal E}_{[V]}=4.38^{+0.62}_{-0.65},
 \label{eq:singlet-enhancement-mc}
\end{equation}
where the interval contains the central $16\%$--$84\%$ range of this
sensitivity calculation.  It is conditional on the equal-weight treatment of
the two alternative production solutions and is not a model-independent
confidence interval.

\subsection{Anomaly--radial-overlap degeneracy}
\label{subsec:anomaly-radial-degeneracy}

For a flavor-singlet radial hybrid $H$, we parameterize the unknown transition
amplitude as
\begin{equation}
 {\cal F}_{H}(0)
 =
 \kappa_A\,R_\gamma\,e^{i\phi_\gamma}\,
 {\cal F}_{\eta_{(2)}}^{[V]}(M_X).
 \label{eq:hybrid-radiative-factorization}
\end{equation}
Here $\kappa_A$ collects the change in the singlet/anomaly coupling between
the lattice $\eta_{(2)}$ and the hybrid, while $R_\gamma$ denotes the
radial and hybrid-operator transition overlap.  The latter can be signed or
complex once mixing and open-channel effects are included.

At the working point, the production input constrains only
\begin{equation}
 \left|\kappa_A R_\gamma\right|=4.38.
 \label{eq:anomaly-radial-product}
\end{equation}
It cannot determine $\kappa_A$ and $R_\gamma$ separately.  In particular,
a large anomaly coupling with a suppressed radial overlap, a moderate anomaly
coupling with an enhanced transition overlap, and additional gluonic or
$c\bar c$ mixing can produce the same radiative rate.

The radiative overlap $R_\gamma$ must also not be identified with the
hadronic-decay node introduced in Sec.~\ref{sec:radial_hybrid_framework}.
The latter is a node in an internal Jacobi momentum entering
$H\to M_1M_2$, whereas $R_\gamma$ is a convolution involving the
$J/\psi$, electromagnetic current, gluonic annihilation kernel, and hybrid
wave function.  The numerical proximity of a photon momentum to a hadronic
breakup momentum would not establish a common zero.

A lattice calculation of $J/\psi\to\gamma\eta_1$ demonstrates that hybrid
production amplitudes with annihilation diagrams can be determined in
principle.  For a $1^{-+}$ hybrid it obtains
$\Gamma=2.29(77)~{\rm eV}$ at a lattice hybrid mass of
$2.23(4)~{\rm GeV}$, and an estimated
${\cal B}(J/\psi\to\gamma\eta_1(1855))=6.2(2.2)\times10^{-5}$
\cite{ChenEtAl2023HybridRadiative}.  We do not use this value as a numerical
anchor for $H(0^{-+})$: the quantum numbers, multipole decomposition, operator
basis, and anomaly coupling are different.

\subsection{Pure-glueball and glueball--\texorpdfstring{$\eta_c$}{eta-c}
benchmarks}
\label{subsec:radiative-mixing-benchmark}

The quenched lattice calculation of pseudoscalar-glueball production obtains
\begin{align}
 m_G &=2.395(14)~{\rm GeV},
 &
 V_G(0)&=0.0246(43),
 &
 {\cal B}(J/\psi\to\gamma G)
 &=2.31(90)\times10^{-4}
 \label{eq:glueball-lattice-anchor}
\end{align}
\cite{GuiEtAl2019PseudoscalarGlueball}.  Holding $V_G(0)$ fixed while moving
the final-state mass by only $36~{\rm MeV}$ to $M_X$ gives
\begin{align}
 {\cal F}_{G}(0;M_X)
 &=0.009018~{\rm GeV}^{-1},
 &
 {\cal B}_{\gamma}^{G}(M_X)
 &=2.67\times10^{-4}.
 \label{eq:pure-glueball-at-x}
\end{align}
The pure-glueball central amplitude is therefore smaller than the working
required amplitude by a factor
\begin{equation}
 \frac{\left|{\cal F}_{X}^{\rm req}\right|}
      {\left|{\cal F}_{G}\right|}
 =3.57.
 \label{eq:pure-glueball-amplitude-deficit}
\end{equation}

A small $c\bar c$ admixture can nevertheless have a large effect because the
connected charmonium transition is much stronger.  Using
\begin{equation}
 V_{c\bar c}(0)=1.865(7)(8)
\end{equation}
gives
\begin{equation}
 {\cal F}_{c\bar c}(0)=1.2268~{\rm GeV}^{-1}
 \label{eq:charmonium-form-factor}
\end{equation}
at the $\eta_c$ mass.  To expose both the sign and possible absorptive phase,
we generalize the usual real two-state convention to
\begin{equation}
 |X\rangle
 =
 \cos\theta\,|G\rangle
 +
 e^{i\delta}\sin\theta\,|c\bar c\rangle ,
 \label{eq:generalized-glueball-charm-mixing}
\end{equation}
so that the benchmark transition amplitude is
\begin{equation}
 {\cal F}_{X}^{\rm mix}(0)
 =
 {\cal F}_{G}(0)\cos\theta
 +
 e^{i\delta}{\cal F}_{c\bar c}(0)\sin\theta .
 \label{eq:mixed-radiative-amplitude}
\end{equation}
The sign convention used in
Ref.~\cite{ChenEtAl2026GlueballEtaCMixing} corresponds to one of the real-phase
limits of Eq.~\eqref{eq:mixed-radiative-amplitude}.

With the central lattice form factors and the constant-${\cal F}_{c\bar c}$
transfer used in this mechanism test, the working production rate is reached at
\begin{align}
 \theta_{\rm cross} &=1.083^\circ,
 && \delta=0,
 \label{eq:constructive-mixing-crossing}
 \\
 \theta_{\rm cross} &=1.925^\circ,
 && \delta=\pi
 \quad\text{after the destructive-interference zero}.
 \label{eq:destructive-mixing-crossing}
\end{align}
Thus an admixture of order one degree can reproduce the required rate, but the
angle inferred from a rate depends strongly on the relative sign.  A continuous
phase produces an equally continuous family of solutions.

These crossings are mechanism benchmarks, not determinations of a physical
mixing angle.  The underlying glueball--charmonium mixing calculation used an
unphysical setup with two degenerate charm sea flavors and no light sea quarks,
and its extrapolation to the physical theory has uncontrolled systematics
\cite{ChenEtAl2026GlueballEtaCMixing}.  Light-quark singlet loops are particularly
important here because the anomaly can evade ordinary OZI suppression and can
alter the magnitude and sign of the mixing contribution.

\subsection{Implementation in the common likelihood}
\label{subsec:radiative-likelihood}

In the common comparison of the radial-hybrid, pure-glueball,
mixed-glueball, and radial-$q\bar q$ hypotheses, the two extracted production
solutions are treated as an alternative-solution mixture:
\begin{equation}
 {\cal L}_{\gamma}(\boldsymbol\Theta_M)
 =
 \frac{1}{2}
 \sum_{s={\rm I,II}}
 {\cal N}\!\left[
 {\cal B}_{\gamma}^{M}(\boldsymbol\Theta_M);
 \widehat{\cal B}_{\gamma,s},
 \sigma_{\gamma,s}^{2}
 +
 \tau_{\gamma}^{2}\widehat{\cal B}_{\gamma,s}^{\,2}
 \right].
 \label{eq:radiative-mixture-likelihood}
\end{equation}
Here $M$ labels the composition hypothesis,
$\boldsymbol\Theta_M$ contains its production parameters, and
$\tau_\gamma$ is a common multiplicative theory nuisance.  The same
$\tau_\gamma$ prior is assigned to every hypothesis.  This construction
avoids treating solutions I and II as independent data and prevents the
radiative block from receiving twice its actual statistical weight.

For the radial-hybrid hypothesis,
\begin{equation}
 {\cal B}_{\gamma}^{H}
 =
 \frac{4\alpha}{27\Gamma_\psi}
 k_{\gamma X}^{3}
 \left|
 \kappa_A R_\gamma
 {\cal F}_{\eta_{(2)}}^{[V]}(M_X)
 \right|^{2}.
 \label{eq:hybrid-production-model}
\end{equation}
For the mixed-glueball hypothesis, the corresponding prediction follows from
Eq.~\eqref{eq:mixed-radiative-amplitude}.  The pure-glueball result is recovered
at $\theta=0$.  A conventional radial-$q\bar q$ model receives the same
production nuisance width, so that its comparison is not artificially
penalized by a narrower form-factor prior.

\begin{figure*}[t]
 \centering
 \includegraphics[width=0.98\textwidth]
 {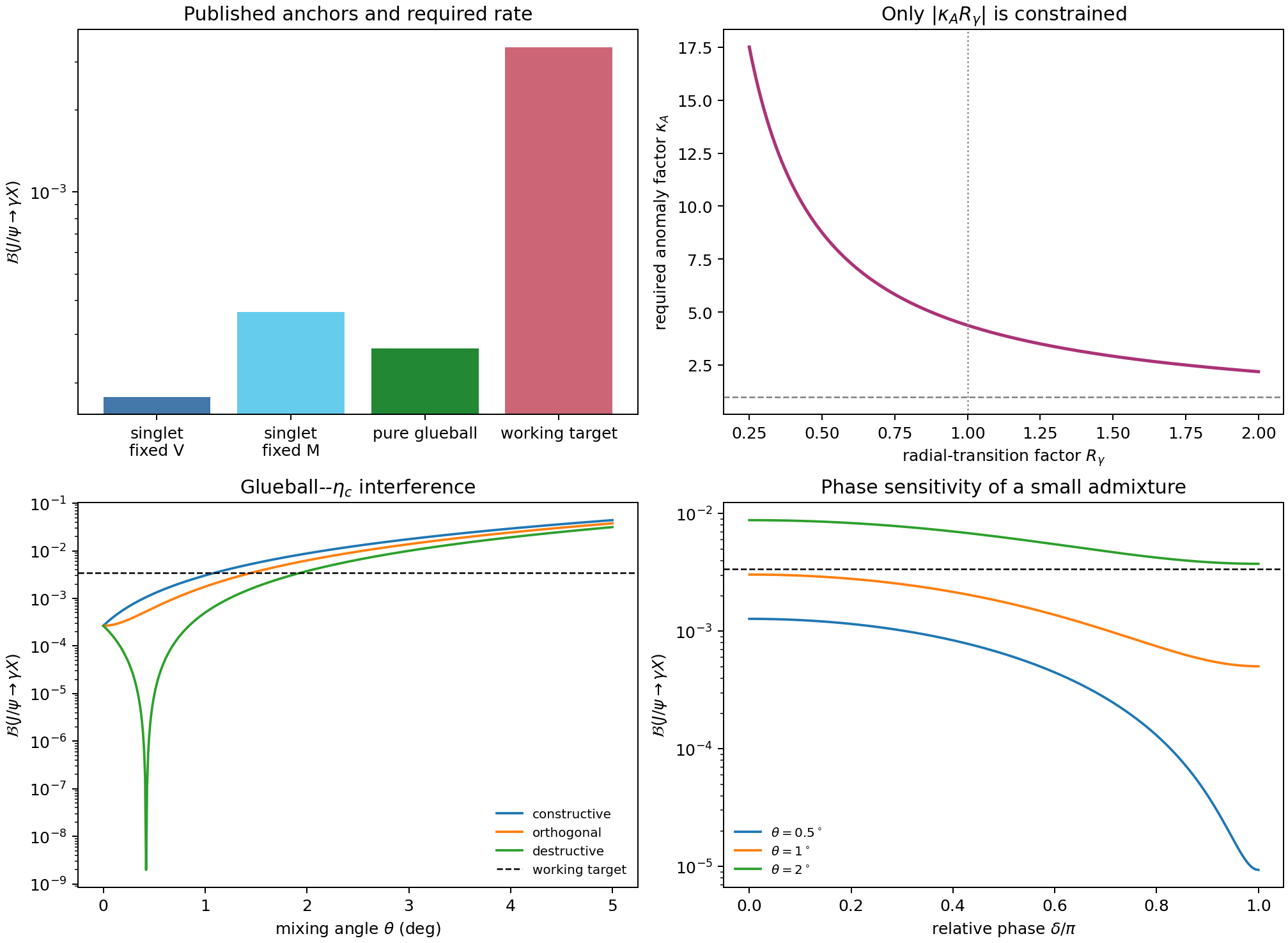}
 \caption{
 Radiative-production diagnostics at $M_X=2.359~{\rm GeV}$.
 Upper left: the two continuations of the $N_f=2$ flavor-singlet lattice
 anchor, the pure-glueball anchor, and the working production rate.
 Upper right: the exact degeneracy between the anomaly factor
 $\kappa_A$ and radial-transition factor $R_\gamma$.
 Lower left: glueball--$\eta_c$ mixing curves for constructive, orthogonal,
 and destructive relative phases.
 Lower right: the phase dependence at three small mixing angles.
 The horizontal dashed line denotes
 ${\cal B}_{\gamma}^{\rm work}=3.4\times10^{-3}$.
 }
 \label{fig:radiative-production-diagnostics}
\end{figure*}

\subsection{Requirements for a dedicated radial-hybrid calculation}
\label{subsec:future-radiative-calculation}

A microscopic prediction of
$J/\psi\to\gamma H(0^{-+})$ should contain substantially more information
than a rescaling of the $\eta_{(2)}$ or $1^{-+}$ lattice result.  At minimum,
it should provide:

\begin{enumerate}
 \item a variational operator basis containing flavor-singlet
 $q\bar q g$, conventional $q\bar q$, and gluonic
 $G\widetilde G$ interpolators, so that radial excitation and configuration
 mixing are determined rather than imposed;

 \item the disconnected light-quark diagrams responsible for the singlet
 anomaly, preferably in $N_f=2+1$ QCD at or near the physical light-quark
 masses;

 \item the three-point function with the charm electromagnetic current over
 sufficient photon virtualities to interpolate to $Q^2=0$, followed by
 continuum and finite-volume extrapolations;

 \item the signed radial-transition overlap and its covariance with the
 extracted state-mixing coefficients, including excited-state contamination
 and open-channel effects; and

 \item a consistent phase convention allowing the hybrid, glueball, and
 $c\bar c$ contributions to be combined coherently.
\end{enumerate}

These requirements are more stringent than reproducing the inclusive
production rate.  A calculation that supplies only
$\left|{\cal F}_{H}(0)\right|$ could test the required magnitude in
Eq.~\eqref{eq:required-form-factor-range}, but it would not by itself determine
whether that magnitude originates from the axial anomaly, radial overlap,
glueball mixing, or a small charmonium component.

\subsection{Radiative-production conclusion}
\label{subsec:radiative-conclusion}

The present radiative data are compatible with a flavor-singlet radial hybrid
provided its effective transition amplitude is approximately $3.1$--$4.4$
times the two direct continuations of the $N_f=2$ singlet anchor.  This is a
required enhancement, not a prediction of the hybrid model.  Conversely, the
pure-glueball lattice anchor lies below the inferred rate, but a charm
admixture of order one degree can bridge the difference, with a substantial
dependence on the relative sign or phase.

Radiative production therefore strengthens the case that the state couples to
flavor-singlet and gluonic dynamics, but it does not distinguish a
flavor-singlet radial hybrid from a mixed-glueball interpretation.  Resolving
that ambiguity requires a dedicated $0^{-+}$ hybrid transition calculation,
a controlled glueball--$q\bar q$--$c\bar c$ mixing analysis, and the
coherent quasi-two-body decay measurements identified in
Sec.~\ref{sec:radial_hybrid_framework}.  The next section consequently compares
all composition hypotheses in one likelihood with common nuisance freedom
rather than using the radiative rate as a standalone discriminator.

\section{Common comparison of the radial-hybrid, glueball, and
\texorpdfstring{$q\bar q$}{qqbar} hypotheses}
\label{sec:common-model-comparison}

The preceding sections show that several properties of $X(2370)$ can be
reproduced by more than one microscopic mechanism.  The suppressed
$K^{*}(892)\bar K$ amplitude can follow from flavor-singlet interference in a
radial hybrid, while large radiative production can follow either from an
enhanced flavor-singlet transition matrix element or from a small
glueball--$\eta_c$ admixture.  Comparisons based on one selected observable
therefore risk attributing a general flavor-singlet or gluonic signature to a
specific constituent assignment.

We instead compare four hypotheses within one prior-predictive likelihood:

\begin{enumerate}
 \item a flavor-singlet first-radial $0^{-+}$ hybrid, denoted $H$;

 \item a pure pseudoscalar-glueball template, denoted $G$;

 \item a glueball--$\eta_c$ mixed template, denoted $G_{\rm mix}$; and

 \item a conventional high radial excitation of a flavor-singlet
 $q\bar q$ pseudoscalar, denoted $Q$.
\end{enumerate}

The objective is not to assign physical posterior probabilities to these four
labels.  The available decay templates do not possess equal microscopic
precision, and the necessary coherent quasi-two-body measurements are not
public.  The calculation is therefore an identifiability stress test: it asks
whether the present data distinguish the hypotheses after they are given
the same generic nuisance-width prescription.

\subsection{Hypothesis definitions and literature anchors}
\label{subsec:model-definitions}

The radial-hybrid decay template is constructed from the signed overlap
calculation in Sec.~\ref{sec:radial_hybrid_framework}, anchored to the
constituent-gluon amplitudes of
Ref.~\cite{FarinaSwanson2024HybridDecays}.  Its mass prior is deliberately
broad enough to cover the current-dependent QCD-sum-rule region
\cite{TanEtAl2024HybridSumRules}, because the proposed state is a radial or
higher hybrid rather than the lowest $0^{-+}$ hybrid.

The pure-glueball production anchor is taken from the lattice pseudoscalar
form factor \cite{GuiEtAl2019PseudoscalarGlueball}, continued over the small
mass difference to $M_X$.  Its decay composition is based on the extended
linear sigma model calculation at $2.37~\mathrm{GeV}$
\cite{EshraimEtAl2013GlueballDecay}.  The mixed-glueball hypothesis uses the
same light-hadron decay template but replaces the radiative-production anchor
by the one-degree constructive glueball--$\eta_c$ benchmark derived in
Sec.~\ref{sec:radiative-production}
\cite{ChenEtAl2026GlueballEtaCMixing}.

The conventional $q\bar q$ hypothesis is retained because high radial
$\eta$- and $\eta'$-like assignments of $X(2370)$ have been studied
within quark-pair-creation models \cite{YuEtAl2011RadialEta}.  Its broad width
and decay-composition inputs are literature-informed surrogates rather than a
new refit of the full radial-$q\bar q$ parameter space.

The central model anchors are summarized in
Table~\ref{tab:common-model-anchors}.

\begin{table*}[t]
 \centering
 \caption{
 Central anchors in the common comparison.  The quoted mass widths specify
 model-predictive envelopes rather than measurements of unmixed basis states.
 The radiative column gives the branching-fraction anchor before the common
 amplitude nuisance is integrated.  The $K^{*}K$ column is the total-mode
 decay-fraction anchor.
 }
 \label{tab:common-model-anchors}
 \begin{tabular}{lcccc}
  \toprule
  Hypothesis
  & Mass anchor [GeV]
  & Width anchor [MeV]
  & ${\cal B}^{(0)}_\gamma$
  & $f^{(0)}_{K^{*}K}$ \\
  \midrule
  Radial hybrid $H$
  & $2.26\pm0.24$
  & $170$
  & $1.7746\times10^{-4}$
  & $1.369\times10^{-4}$ \\
  Pure glueball $G$
  & $2.60\pm0.30$
  & $170$
  & $2.6662\times10^{-4}$
  & $5.0\times10^{-3}$ \\
  Mixed glueball $G_{\rm mix}$
  & $2.42\pm0.22$
  & $177$
  & $3.0355\times10^{-3}$
  & $5.0\times10^{-3}$ \\
  Radial $q\bar q$, $Q$
  & $2.37\pm0.14$
  & $350$
  & $2.4800\times10^{-4}$
  & $8.0\times10^{-2}$ \\
  \bottomrule
 \end{tabular}
\end{table*}

The equal total-width anchors assigned to $H$ and $G$ are intentional.
Neither available calculation provides a complete, equally controlled
absolute-width prediction including every open channel at $2.359~\mathrm{GeV}$.
Their width anchors are consequently centered on the observed scale and given
the same multiplicative uncertainty.  The total-width block is therefore
expected to carry little hybrid--glueball discrimination.  It is retained to
test the much broader conventional radial-$q\bar q$ template.

For the inclusive three-pseudoscalar composition, we use the channel ordering
\begin{equation}
 \boldsymbol f_{\rm PPP}
 =
 \left(
 f_{K\bar K\pi},
 f_{\pi\pi\eta},
 f_{\pi\pi\eta'},
 f_{K\bar K\eta'}
 \right).
 \label{eq:ppp-channel-order}
\end{equation}
The normalized central templates are
\begin{subequations}
\label{eq:ppp-model-templates}
\begin{align}
 \boldsymbol t_H
 &=
 (0.510,\;0.299,\;0.130,\;0.061),
 \label{eq:hybrid-ppp-template}
 \\
 \boldsymbol t_G
 =
 \boldsymbol t_{G_{\rm mix}}
 &=
 (0.640,\;0.223,\;0.122,\;0.015),
 \label{eq:glueball-ppp-template}
 \\
 \boldsymbol t_Q
 &=
 (0.400,\;0.150,\;0.200,\;0.250).
 \label{eq:qqbar-ppp-template}
\end{align}
\end{subequations}
The equality
$\boldsymbol t_G=\boldsymbol t_{G_{\rm mix}}$ is a limitation, not a physical
prediction that mixing leaves every decay amplitude unchanged.  Current
decay-composition data can distinguish the hybrid template from this shared
glueball template, but cannot separate pure from mixed glueball through this
block alone.

The hybrid PPP vector is a phenomenological handoff from the two-body
template, not a microscopic three-body calculation.  Before normalization we
map the modeled weights as
\begin{align}
 \boldsymbol w_H^{\rm PPP}&=\bigl(
 w_{K_0^*K}+w_{K^*K},\;
 w_{a_0\pi}+w_{f_0(500)\eta}+w_{f_0(980)\eta},\nonumber\\
 &\hspace{2.6cm}0.45[w_{f_0(500)\eta}+w_{f_0(980)\eta}],\;
 0.12[w_{K_0^*K}+w_{K^*K}]\bigr),\nonumber\\
 \boldsymbol t_H&=\boldsymbol w_H^{\rm PPP}/\sum_iw_{H,i}^{\rm PPP}.
 \label{eq:hybrid-ppp-handoff}
\end{align}
The coefficients $0.45$ and $0.12$ encode unresolved $\eta$--$\eta'$,
charge, and subchannel feed-through.  They are fixed surrogate inputs; hence
the PPP score is conditional on this mapping and is not a prediction of a
unitary Dalitz-plot model.

For reproducibility, the glueball vector is the normalization of
$(0.46,0.16,0.088,0.011)$, assembled from the $2.37~\mathrm{GeV}$ eLSM
three-pseudoscalar pattern.  The $Q$ vector is an explicit phenomenological
surrogate, not a number quoted by one reference.  Likewise, the glueball proxy
weights $(0.033,0.100,0.069)$ use $0.100$ and $0.069$ as eLSM-inspired
$a_0\pi$ and $K_0^*K$ anchors, while $0.033$ is assigned to the light-scalar
$f_0(980)\eta$ proxy; the $Q$ values are model-design inputs.  These fixed
central mappings and the hybrid coefficients above are not marginalized
separately.  The common $\sigma_d$ covariance permits broad component-wise
deformations, but the resulting scores remain conditional on the stated
central handoffs.

\subsection{Current data vector and missing-data gate}
\label{subsec:common-data-vector}

The current-data vector is divided into four blocks,
\begin{equation}
 {\cal D}_{\rm now}
 =
 \left\{
 {\cal D}_m,\,
 {\cal D}_\Gamma,\,
 {\cal D}_{\gamma,K^{*}K},\,
 {\cal D}_{\rm PPP}
 \right\}.
 \label{eq:current-data-vector}
\end{equation}
These contain, respectively, the measured mass, total width, phenomenologically
inferred radiative-production solutions together with the direct visible
$K^{*}K$ product result, and the normalized inclusive three-pseudoscalar
composition.

The $K^{*}K$ input is kept in its directly measured visible form,
\begin{equation}
 \widehat{\cal P}_{K^{*}K}^{\,\rm vis}
 =
 (0.1\pm1.2_{\rm stat}\pm1.1_{\rm syst})\times10^{-6},
 \label{eq:visible-kstar-input}
\end{equation}
with the reported $90\%$ confidence-level upper limit
$2.7\times10^{-6}$
\cite{BESIII2026X2370Glueball}.  The predicted visible product is
\begin{equation}
 {\cal P}_{K^{*}K}^{\,\rm vis}
 =
 {\cal B}_{\gamma}\,
 f_{K^{*}K}\,
 c_{\rm vis},
 \qquad
 c_{\rm vis}=\frac{1}{6},
 \label{eq:visible-kstar-prediction}
\end{equation}
where $c_{\rm vis}$ contains the neutral-cascade and charge/isospin
conversion adopted in the experimental comparison.

The requested coherent fractions,
\begin{equation}
 \boldsymbol f_{\rm q2b}
 =
 \left(
 f_{f_0(980)\eta},
 f_{a_0(980)\pi},
 f_{K_0^{*}(1430)K}
 \right),
 \label{eq:q2b-fraction-vector}
\end{equation}
and their covariance matrix have not been released publicly.  The likelihood
therefore enforces the missing-data gate
\begin{equation}
 {\cal L}_{\rm q2b}({\cal D}_{\rm now}\mid M)\equiv1.
 \label{eq:q2b-missing-data-gate}
\end{equation}
No synthetic fraction, Asimov value, or independently fitted resonance yield
is admitted to the current-data score.  Forecast calculations using
Eq.~\eqref{eq:q2b-fraction-vector} are reported separately below.

\subsection{Factorized likelihood}
\label{subsec:common-likelihood}

For hypothesis $M$, the screening likelihood is factorized as
\begin{equation}
 {\cal L}({\cal D}_{\rm now}\mid\boldsymbol\theta_M,M)
 =
 {\cal L}_{m}\,
 {\cal L}_{\Gamma}\,
 {\cal L}_{\gamma,K^{*}K}\,
 {\cal L}_{\rm PPP}.
 \label{eq:factorized-common-likelihood}
\end{equation}
This factorization permits an exact observable-block decomposition of the
score.  The production and visible-$K^{*}K$ terms remain in one joint
block because they share the same latent radiative branching fraction.

For the model mass anchor $\mu_{m,M}\pm s_{m,M}$, we preserve the measured
mass asymmetry through
\begin{equation}
 {\cal L}_m
 =
 p_{\rm SN}\!\left(
 \widehat M_X\mid \mu_{m,M},
 \sqrt{(\sigma_M^-)^2+r_m^2s_{m,M}^2},
 \sqrt{(\sigma_M^+)^2+r_m^2s_{m,M}^2}
 \right),
 \label{eq:mass-likelihood}
\end{equation}
where $r_m$ is a common model-mass uncertainty multiplier.  The quadrature
addition of the model envelope to the two experimental sides is an explicit
approximation to a numerical Gaussian--split-normal convolution.

The total width is assigned a log-normal model nuisance,
\begin{equation}
 \Gamma_M
 =
 \Gamma_M^{(0)}
 \exp\!\left(\sigma_\Gamma z_\Gamma\right),
 \qquad
 z_\Gamma\sim{\cal N}(0,1),
 \label{eq:width-nuisance}
\end{equation}
and is compared as $p_{\rm SN}(\widehat\Gamma_X\mid\Gamma_M,
\sigma_\Gamma^-,\sigma_\Gamma^+)$, so that the side is selected by the
observed residual relative to the prediction.

For production, the common amplitude-level nuisance is
\begin{equation}
 {\cal B}_{\gamma,M}
 =
 {\cal B}_{\gamma,M}^{(0)}
 \exp\!\left(2\sigma_A z_A\right),
 \qquad
 z_A\sim{\cal N}(0,1).
 \label{eq:production-nuisance}
\end{equation}
The factor of two follows because the branching fraction is quadratic in the
transition amplitude.  The two phenomenological production solutions are
treated as alternative descriptions,
\begin{equation}
 {\cal L}_{\gamma}
 =
 \frac{1}{2}
 \sum_{s={\rm I,II}}
 {\cal N}\!\left(
 {\cal B}_{\gamma,M};
 \widehat{\cal B}_{\gamma,s},
 \sigma_{\gamma,s}^{2}
 \right),
 \label{eq:production-solution-mixture}
\end{equation}
rather than as two independent measurements
\cite{SunEtAl2022NatureX2370}.

The $K^{*}K$ fraction receives the common multiplicative decay nuisance
\begin{equation}
 f_{K^{*}K,M}
 =
 f_{K^{*}K,M}^{(0)}
 \exp\!\left(\sigma_d z_{K^{*}K}\right),
 \qquad
 z_{K^{*}K}\sim{\cal N}(0,1),
 \label{eq:kstar-fraction-nuisance}
\end{equation}
with the physical range enforced numerically.  Combining
Eqs.~\eqref{eq:visible-kstar-prediction},
\eqref{eq:production-nuisance}, and
\eqref{eq:kstar-fraction-nuisance} defines
${\cal L}_{\gamma,K^{*}K}$.

Because the four inclusive PPP components form a composition, their
uncertainties are represented in additive-log-ratio coordinates,
\begin{equation}
 y_i
 =
 \log\frac{f_i}{f_4},
 \qquad i=1,2,3.
 \label{eq:alr-coordinates}
\end{equation}
If $\boldsymbol\mu_{\rm ALR}$ and
$\boldsymbol C_{\rm ALR}^{\rm exp}$ denote the propagated experimental mean
and covariance, the model likelihood is
\begin{equation}
 {\cal L}_{\rm PPP}
 =
 {\cal N}_3\!\left[
 \boldsymbol\mu_{\rm ALR};
 \boldsymbol y(\boldsymbol t_M),
 \boldsymbol C_{\rm ALR}^{\rm exp}
 +\sigma_d^2
 \left(
 \boldsymbol I+\boldsymbol 1\boldsymbol 1^{T}
 \right)
 \right].
 \label{eq:ppp-alr-likelihood}
\end{equation}
The additional covariance is generated by independent common-width
log-normal perturbations of the four template components.
For the headline comparison,
$\boldsymbol\mu_{\rm ALR}=\langle\boldsymbol y\rangle$ and
$\boldsymbol C_{\rm ALR}^{\rm exp}=\boldsymbol C_y$ are taken directly from
Eqs.~\eqref{eq:benchmark_alr_mean} and
\eqref{eq:benchmark_alr_covariance}, corresponding to
$\rho_{\rm sys}=0.5$.  The covariance scan changes only $\rho_{\rm sys}$;
all model anchors, likelihood blocks, nuisance distributions, and quadrature
settings remain fixed.

\subsection{Common nuisance scenarios and screening scores}
\label{subsec:common-nuisance-scenarios}

Every hypothesis is assigned the same nuisance widths.  The three principal
settings are given in Table~\ref{tab:nuisance-scenarios}.

\begin{table}[t]
 \centering
 \caption{
 Common nuisance settings.  The first three columns are standard deviations
 in logarithmic coordinates.  The same values are used for every hypothesis.
 }
 \label{tab:nuisance-scenarios}
 \begin{tabular}{lcccc}
  \toprule
  Scenario
  & $\sigma_A$
  & $\sigma_d$
  & $\sigma_\Gamma$
  & $r_m$ \\
  \midrule
  Tight    & $0.45$ & $0.45$ & $0.35$ & $1.0$ \\
  Baseline & $0.80$ & $0.80$ & $0.55$ & $1.0$ \\
  Wide     & $1.20$ & $1.20$ & $0.80$ & $1.5$ \\
  \bottomrule
 \end{tabular}
\end{table}

For compactness we define the conditional integrated screening score
\begin{equation}
 Z_M
 =
 \int d\boldsymbol\theta_M\,
 \pi(\boldsymbol\theta_M\mid M)\,
 {\cal L}({\cal D}_{\rm now}\mid\boldsymbol\theta_M,M).
 \label{eq:model-screening-score}
\end{equation}
The mass and inclusive-composition integrals are evaluated analytically.
The width and joint production--$K^{*}K$ integrals are evaluated by
deterministic adaptive quadrature over standard-normal nuisance variables.
Doubling the numerical quadrature order from $48$ to $96$ leaves all
reported hybrid--mixed-glueball differences unchanged at the displayed
precision.

Several anchors are phenomenological surrogates, and the $H$ and $G$ width
centers are set to the observed width scale.  Thus $Z_M$ is not a Bayesian
evidence based wholly on external priors: the width block measures relative
tolerance around a common empirical scale and cannot validate either model.
We report
\begin{equation}
 \Delta\log Z_M
 =
 \log Z_M-\max_{M'}\log Z_{M'},
 \label{eq:delta-log-evidence}
\end{equation}
within each nuisance scenario.  Because the model templates are not equally
microscopic, these score differences are conditional comparisons of the
implemented hypotheses.  They are not universal Bayes factors between all
possible hybrid, glueball, and $q\bar q$ models.

\subsection{Present-data comparison}
\label{subsec:current-model-comparison-results}

The resulting score differences are given in
Table~\ref{tab:current-model-comparison}.

\begin{table}[t]
 \centering
 \caption{
 Integrated screening score relative to the best model in each nuisance scenario, using
 the benchmark asymmetric PPP covariance with $\rho_{\rm sys}=0.5$.
 Values close to zero indicate that the available data do not resolve the
 hypotheses at the precision of the implemented templates.
 }
 \label{tab:current-model-comparison}
 \begin{tabular}{lrrr}
  \toprule
  Hypothesis
  & Tight
  & Baseline
  & Wide \\
  \midrule
  Radial hybrid $H$
  & $-1.289$
  & $0.000$
  & $0.000$ \\
  Pure glueball $G$
  & $-3.288$
  & $-1.614$
  & $-1.309$ \\
  Mixed glueball $G_{\rm mix}$
  & $0.000$
  & $-0.035$
  & $-0.415$ \\
  Radial $q\bar q$, $Q$
  & $-12.623$
  & $-8.046$
  & $-5.132$ \\
  \bottomrule
 \end{tabular}
\end{table}

The central result is
\begin{equation}
 \left.
 \log\frac{Z_H}{Z_{G_{\rm mix}}}
 \right|_{\rm baseline,\;\rho_{\rm sys}=0.5}
 =
 +0.0350.
 \label{eq:baseline-hybrid-mixed-evidence}
\end{equation}
The corresponding conditional score ratio is $1.04$, effectively unity at
the precision of these screening templates.  The tight scenario favors the
mixed-glueball template, whereas the wide scenario favors the radial hybrid:
\begin{align}
 \left.
 \log\frac{Z_H}{Z_{G_{\rm mix}}}
 \right|_{\rm tight}
 &=-1.289,
 &
 \left.
 \log\frac{Z_H}{Z_{G_{\rm mix}}}
 \right|_{\rm wide}
 &=+0.415.
 \label{eq:hybrid-mixed-scenario-dependence}
\end{align}
The exchange of ordering is more informative than the identity of the winner
in any single column.

The conventional radial-$q\bar q$ surrogate is below the best model by
$5.13$--$12.62$ log-score units.  This conditional result is driven by its broad
width, larger $K^{*}K$ fraction, and different PPP template.  It should not be
reported as a model-independent exclusion of all radial-$q\bar q$ states:
radial nodes, mixing angles, coupled-channel dressing, and alternative decay
parameters have not been exhaustively refitted.

\subsection{Observable-block decomposition}
\label{subsec:evidence-decomposition}

Because Eq.~\eqref{eq:factorized-common-likelihood} factorizes, the
hybrid--mixed-glueball score difference is exactly
\begin{equation}
 \log\frac{Z_H}{Z_{G_{\rm mix}}}
 =
 \sum_b
 \left(
 \log Z_{H,b}
 -
 \log Z_{G_{\rm mix},b}
 \right),
 \label{eq:block-evidence-sum}
\end{equation}
where $b$ runs over mass, width, joint production--$K^{*}K$, and inclusive
PPP composition.  At baseline,
\begin{align}
 \Delta_m
 &=-0.1333,
 &
 \Delta_\Gamma
 &=+0.0178,
 \nonumber\\
 \Delta_{\gamma,K^{*}K}
 &=-0.5594,
 &
 \Delta_{\rm PPP}
 &=+0.7099.
 \label{eq:baseline-block-contributions}
\end{align}
Their sum is
\begin{equation}
 -0.1333+0.0178-0.5594+0.7099=+0.0350.
 \label{eq:block-contribution-sum-numeric}
\end{equation}

The mass and total width provide little discrimination.  The near-tie instead
results from cancellation between two moderately informative blocks:
radiative production together with the $K^{*}K$ constraint favors the
mixed-glueball template, whereas inclusive three-pseudoscalar composition
favors the radial hybrid.  Removing either block reverses or strengthens the
ordering.  The comparison is therefore not a redundant accumulation of
glueball-like observables.

\subsection{Continuous common-prior sensitivity}
\label{subsec:continuous-prior-scan}

To test whether the three discrete scenarios conceal a sharp prior boundary,
we introduce a continuous common rescaling parameter $\lambda$:
\begin{equation}
 \sigma_A=0.80\lambda,
 \qquad
 \sigma_d=0.80\lambda,
 \qquad
 \sigma_\Gamma=0.55\lambda,
 \label{eq:continuous-nuisance-scaling}
\end{equation}
while holding the mass-envelope multiplier at its baseline value.  The same
$\lambda$ is applied to every hypothesis.  Scanning
$0.5\leq\lambda\leq1.5$ gives
\begin{equation}
 \log Z_H-\log Z_{G_{\rm mix}}=0
 \quad\text{at}\quad
 \lambda_{\rm cross}=0.9747.
 \label{eq:continuous-prior-crossing}
\end{equation}
Thus narrowing the baseline amplitude, decay, and width nuisance scales by
only about $2.5\%$ reverses the ordering.

This crossing is not used to select a preferred prior after examining the
data.  It is a robustness diagnostic demonstrating that the present
hybrid--mixed-glueball ranking is controlled by nuisance-volume assumptions.
Normalized score weights obtained after assigning equal prior weight
to the four implemented templates must consequently not be interpreted as
measured constituent probabilities.

\subsection{Conditional production-anchor sensitivity}
\label{subsec:production-anchor-sensitivity}

The radial-hybrid production anchor in
Table~\ref{tab:common-model-anchors} is the fixed-$V(0)$ continuation of the
$N_f=2$ singlet result.  To expose its impact, we replace it conditionally by
\begin{equation}
 {\cal B}_{\gamma,H}^{(0)}(r_H)
 =
 r_H^2\,
 {\cal B}_{\gamma,H}^{(0)}(r_H=1),
 \label{eq:hybrid-anchor-profile}
\end{equation}
where $r_H$ is the effective anomaly--radial amplitude factor introduced in
Sec.~\ref{sec:radiative-production}.  Retaining the baseline common nuisance
widths, the hybrid score is maximized on the scan at
\begin{equation}
 r_H^{\rm best}=4.04.
 \label{eq:hybrid-anchor-best-factor}
\end{equation}
This is close to the independent matrix-element requirement $4.38$ found from
the working radiative rate.  At the grid optimum,
\begin{equation}
 \log Z_H-\log Z_{G_{\rm mix}}=+1.54.
 \label{eq:hybrid-profiled-evidence}
\end{equation}
This does not establish that the physical enhancement equals four.
It shows that a future microscopic calculation predicting an enhancement of
this magnitude would remove the current radiative-production penalty against
the hybrid.

For the mixed-glueball hypothesis, the conditional amplitude is
\begin{equation}
 {\cal F}_{G_{\rm mix}}(0)
 =
 {\cal F}_{G}(0)\cos\theta
 +
 e^{i\delta}
 {\cal F}_{c\bar c}(0)\sin\theta .
 \label{eq:mixed-anchor-profile}
\end{equation}
The best grid angles are
\begin{align}
 \theta_{\rm best}
 &=
 0.9^\circ,
 &&\delta=0,
 \nonumber\\
 &=
 1.3^\circ,
 &&\delta=\frac{\pi}{2},
 \nonumber\\
 &=
 1.8^\circ,
 &&\delta=\pi.
 \label{eq:mixed-profile-best-angles}
\end{align}
All three phase choices can reproduce nearly the same optimum score.
Near the destructive-interference zero, however, the mixed-glueball production
anchor becomes strongly suppressed and the hybrid is favored.  Consequently,
a production rate alone cannot determine a unique mixing angle without a
signed transition calculation and a normalized physical mixing prior.

The scans in Eqs.~\eqref{eq:hybrid-anchor-profile} and
\eqref{eq:mixed-anchor-profile} are conditional profiles, not marginal
posteriors.  Promoting them to external-prior model evidences would require normalized priors
for $r_H$, $\theta$, and $\delta$, derived independently of the
$X(2370)$ production data.

\subsection{Forecast for coherent quasi-two-body fractions}
\label{subsec:q2b-forecast}

Although the observables in Eq.~\eqref{eq:q2b-fraction-vector} are unavailable,
their expected discriminating power can be explored without inserting them
into the current score.  The numbers below are normalized two-body
intensity proxies.  They are not yet experimental coherent fit fractions,
whose denominators contain interference and whose sum need not equal unity.
The central proxy templates are
\begin{equation}
\begin{array}{c|ccc}
 & f_{f_0(980)\eta}
 & f_{a_0(980)\pi}
 & f_{K_0^{*}(1430)K} \\
\hline
 H
 &0.0925&0.00442&0.240\\
 G,\;G_{\rm mix}
 &0.033&0.100&0.069\\
 Q
 &0.015&0.035&0.040
\end{array}.
\label{eq:q2b-central-templates}
\end{equation}

For hybrid truth, the forecast statistic for candidate $M$ is
\begin{equation}
 \Delta\chi^2_{H\to M}
 =
 \left(
 \boldsymbol f_M-\boldsymbol f_H
 \right)^T
 \left(
 \boldsymbol C_{\rm exp}
 +\boldsymbol C_{\rm th}^{H}
 +
 \boldsymbol C_{\rm th}^{M}
 \right)^{-1}
 \left(
 \boldsymbol f_M-\boldsymbol f_H
 \right).
 \label{eq:q2b-asimov-statistic}
\end{equation}
We define
\begin{equation}
 C_{{\rm th},ij}^{M}
 =
 \delta_{ij}
 \left(
 \tau_{\rm th}f_{M,i}
 \right)^2,
 \label{eq:q2b-theory-covariance}
\end{equation}
and construct $\boldsymbol C_{\rm exp}$ from a common marginal fractional
uncertainty with pairwise correlation $0.3$.  At the benchmark point
\begin{equation}
 \frac{\sigma_i^{\rm exp}}{f_{H,i}}=0.20,
 \qquad
 \tau_{\rm th}=0.50,
 \label{eq:q2b-forecast-benchmark}
\end{equation}
we obtain
\begin{align}
 \Delta\chi^2_{H\to G}
 =
 \Delta\chi^2_{H\to G_{\rm mix}}
 &=6.52,
 \label{eq:q2b-forecast-glueball}
 \\
 \Delta\chi^2_{H\to Q}
 &=7.61.
 \label{eq:q2b-forecast-qqbar}
\end{align}
These values reproduce the earlier fixed benchmark and remain separated from
the current-data likelihood.

The equality in Eq.~\eqref{eq:q2b-forecast-glueball} follows because the pure
and mixed glueball hypotheses share the same decay template.  Coherent decay
fractions can therefore distinguish the radial-hybrid pattern from the shared
glueball pattern, but an additional production or mixing-sensitive observable
is required to distinguish $G$ from $G_{\rm mix}$.

The forecast has also been scanned over
\begin{equation}
 0.10\leq
 \frac{\sigma_i^{\rm exp}}{f_{H,i}}
 \leq0.50,
 \qquad
 0.25\leq\tau_{\rm th}\leq1.00.
 \label{eq:q2b-forecast-grid}
\end{equation}
These quantities are Asimov distances between uncertain proxy templates.  A
real-data test must forward-model the predicted complex amplitudes through the
experimental line shapes, acceptance, and interference convention.  The
numbers should therefore not be treated as a ready-made likelihood or
converted directly into Gaussian significances without a
complete amplitude likelihood, physical boundaries, interference fractions,
and a validated theory covariance.

\begin{figure*}[t]
 \centering
 \includegraphics[width=0.98\textwidth]
 {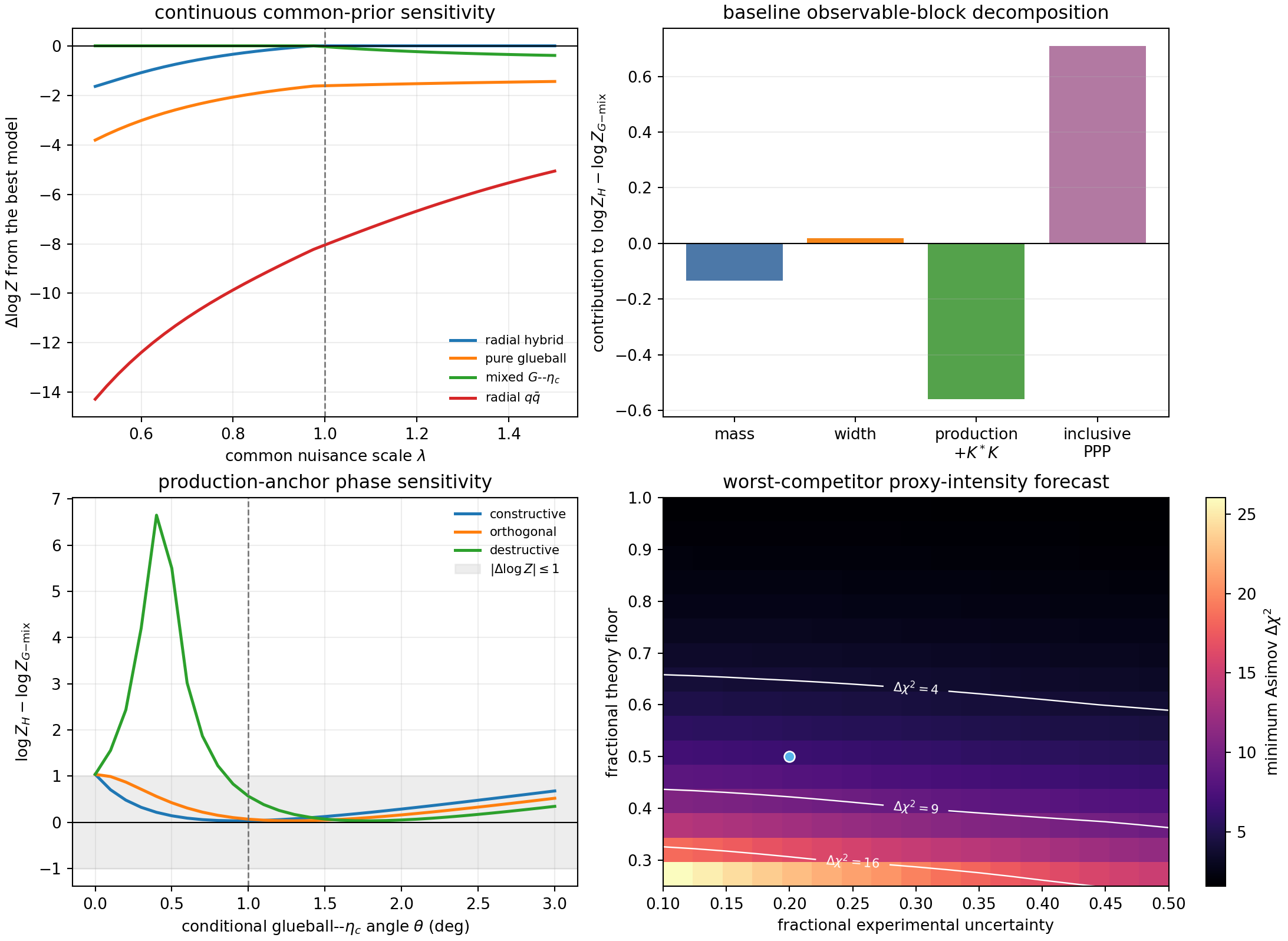}
 \caption{
 Common model-comparison diagnostics.
 Upper left: screening score relative to the best hypothesis as the baseline
 non-mass nuisance widths are multiplied by a common scale $\lambda$.
 Upper right: exact baseline contributions to
 $\log Z_H-\log Z_{G_{\rm mix}}$.
 Lower left: conditional hybrid--mixed score difference as a function of
 the glueball--$\eta_c$ mixing angle and relative phase; the shaded region
 denotes $|\Delta\log Z|\leq1$.
 Lower right: minimum hybrid-truth Asimov $\Delta\chi^2$ against the
 competing decay templates as the experimental precision and common theory
 floor are varied.  The blue point denotes the $20\%$, $50\%$ benchmark.
 Forecast quantities are not included in the present-data score.
 }
 \label{fig:common-model-comparison}
\end{figure*}

\subsection{Interpretation and scope}
\label{subsec:model-comparison-interpretation}

The common analysis leads to four conclusions.

First, the flavor-singlet radial-hybrid and mixed-glueball hypotheses are
indistinguishable with the current data and implemented theory precision.
Their baseline score ratio is essentially unity, and a small common
rescaling of the nuisance widths changes their ordering.

Second, radiative production and inclusive hadronic composition pull in
opposite directions.  The former favors the mixed-glueball production
mechanism, whereas the latter favors the radial-hybrid decay template.  The
mass and total width do not resolve the competition.

Third, the conventional radial-$q\bar q$ surrogate is disfavored within the
implemented template family, but this is not a general exclusion of
conventional pseudoscalar excitations.  A publication-level exclusion would
require refitting their radial wave functions, flavor mixing, decay operator,
and coupled-channel dressing with nuisance freedom comparable to that of the
hybrid and glueball calculations.

Fourth, coherent measurements of
$f_0(980)\eta$ or $a_0(980)\pi$, together with
$K_0^{*}(1430)K$, can distinguish the hybrid decay pattern from the shared
glueball pattern over a broad range of experimental and theoretical
uncertainties.  They cannot by themselves distinguish pure from mixed glueball.
That final separation requires a direct radiative-production measurement,
a signed $J/\psi\to\gamma H(0^{-+})$ calculation, or an independent
mixing-sensitive observable.

Accordingly, the present result supports the statement that a flavor-singlet
radial hybrid is compatible with current data, but remains experimentally and
theoretically indistinguishable from a mixed-glueball interpretation.  The
comparison does not support a claim that either component has been uniquely
identified.

\section{Coherent-amplitude measurements that could help resolve the ambiguity}
\label{sec:experimental-discrimination}

The common comparison developed above establishes an identifiability problem:
with the presently available mass, total-width, radiative-production,
inclusive three-pseudoscalar, and visible $K^{*}(892)K$ information, the
flavor-singlet radial-hybrid and mixed-glueball hypotheses remain
indistinguishable under reasonable common nuisance freedom.  We now ask which
additional measurements would most efficiently break this degeneracy.  The
analysis in this section is an Asimov measurement-design study.  No synthetic
fraction or covariance matrix is included in the current-data likelihood.

\subsection{Missing coherent information}
\label{subsec:missing-coherent-information}

The observation of an $a_{0}(980)^{0}\pi^{0}$ contribution with a significance
above $9\sigma$ in
$J/\psi\to\gamma\pi^{0}\pi^{0}\eta$ demonstrates that relevant
scalar--pseudoscalar substructure is experimentally accessible
\cite{BESIII2026X2370NeutralModes}.  However, the public analyses do not provide the common
three-component vector
\begin{equation}
 \boldsymbol f =
 \left(
  f_{f_{0}(980)\eta},
  f_{a_{0}(980)\pi},
  f_{K_{0}^{*}(1430)K}
 \right),
 \label{eq:design-q2b-fraction-vector}
\end{equation}
together with its statistical and systematic covariance.  Observation
significance or an independently fitted event yield is not equivalent to such
a measurement.

For an amplitude model
\begin{equation}
 {\cal A}(\Phi)=\sum_i c_i {\cal A}_i(\Phi),
 \label{eq:coherent-amplitude-sum}
\end{equation}
a diagonal fit fraction is conventionally defined as
\begin{equation}
 F_i =
 \frac{
  \displaystyle\int d\Phi\,
  \left|c_i{\cal A}_i(\Phi)\right|^2
 }{
  \displaystyle\int d\Phi\,
  \left|\sum_j c_j{\cal A}_j(\Phi)\right|^2
 },
 \label{eq:fit-fraction-definition}
\end{equation}
whereas the integrated interference fraction between two components is
\begin{equation}
 F_{ij} =
 \frac{
  \displaystyle
  2\,\mathrm{Re}\!
  \int d\Phi\,
  c_i c_j^{*}{\cal A}_i(\Phi){\cal A}_j^{*}(\Phi)
 }{
  \displaystyle\int d\Phi\,
  \left|\sum_k c_k{\cal A}_k(\Phi)\right|^2
 }.
 \label{eq:interference-fraction-definition}
\end{equation}
Consequently,
\begin{equation}
 \sum_i F_i+\sum_{i<j}F_{ij}=1,
 \qquad
 \sum_i F_i\neq 1
 \quad\text{in general}.
 \label{eq:fit-fraction-sum}
\end{equation}
Fit fractions therefore cannot be treated as independent probabilities or as
an ordinary composition vector.  The preferred reusable experimental object
is the complex coefficient vector $c_i$, or the experimental likelihood for
that vector, with a stated phase convention and complete covariance
\cite{BackEtAl2018Laura,AlbaladejoEtAl2022AmplitudeAnalysis}.

\subsection{Forecast definition}
\label{subsec:measurement-forecast-definition}

For direct continuity with the common model comparison, we use the
quasi-two-body templates
\begin{align}
 \boldsymbol f_H &=
 (0.09254,\;0.004425,\;0.24015), \nonumber\\
 \boldsymbol f_G &=
 (0.033,\;0.100,\;0.069), \nonumber\\
 \boldsymbol f_Q &=
 (0.015,\;0.035,\;0.040),
 \label{eq:measurement-design-templates}
\end{align}
where $H$, $G$, and $Q$ denote the radial-hybrid, glueball, and
conventional radial-$q\bar q$ templates, respectively.  The pure- and
mixed-glueball hypotheses share $\boldsymbol f_G$ in the present
implementation.  Thus, the observables considered here can separate the
hybrid decay pattern from the common glueball pattern, but cannot by themselves
distinguish a pure glueball from a glueball--$\eta_c$ mixed state.

For a subset $S$ of the three channels, the hybrid-truth Asimov separation
from competitor $M\in\{G,Q\}$ is
\begin{equation}
 \Delta\chi^2_{H\rightarrow M}(S)=
 \left(
  \boldsymbol f_{M,S}-\boldsymbol f_{H,S}
 \right)^{T}
 C^{-1}_{M,S}
 \left(
  \boldsymbol f_{M,S}-\boldsymbol f_{H,S}
 \right).
 \label{eq:subset-asimov-dchi2}
\end{equation}
The forecast covariance is
\begin{align}
 C_{M,S}
 &=
 C_{\mathrm{exp},S}+C_{\mathrm{th},H,S}+C_{\mathrm{th},M,S},
 \label{eq:forecast-total-covariance}\\
 C_{\mathrm{exp},S}
 &=
 R_{\rho,S}\circ
 \left(\boldsymbol\sigma_S\boldsymbol\sigma_S^{T}\right),
 \qquad
 \sigma_i=\max\!\left(r f_{H,i},f_{\mathrm{abs}}\right),
 \label{eq:forecast-experimental-covariance}\\
 C_{\mathrm{th},M,S}
 &=
 \operatorname{diag}
 \left[
  \left(\tau f_{M,i}\right)^2
 \right]_{i\in S},
 \label{eq:forecast-theory-covariance}
\end{align}
where $\circ$ is the Hadamard product.  The benchmark values are
\begin{equation}
 r=0.20,\qquad
 \rho=0.30,\qquad
 \tau=0.50,\qquad
 f_{\mathrm{abs}}=0.005.
 \label{eq:measurement-design-benchmark}
\end{equation}
Here $r$ is the marginal relative experimental uncertainty,
$R_{\rho}$ is an equicorrelation matrix, and $\tau$ is a benchmark fractional
theory floor.  The absolute floor prevents the very
small hybrid $a_{0}(980)\pi$ fraction from receiving unrealistically high
relative precision.  These quantities specify a sensitivity benchmark, not a
reconstruction of an unpublished experimental covariance.  The diagonal
theory model neglects normalization-induced and shared-model correlations; a
microscopic correlated theory covariance could either increase or decrease the
reported separation.

\begin{table*}[t]
 \centering
 \caption{
 Benchmark hybrid-truth Asimov separations for all nonempty channel subsets.
 The glueball column applies to both the pure- and mixed-glueball hypotheses
 because they share the same decay template.  The final column is the
 worst-competitor sensitivity for each subset.  These are forecasts, not
 present-data test statistics.
 }
 \label{tab:channel-subset-forecast}
 \begin{tabular}{lccc}
  \toprule
  Channel subset $S$
   & $\Delta\chi^2_{H\rightarrow G}$
   & $\Delta\chi^2_{H\rightarrow Q}$
   & $\min_M\Delta\chi^2_{H\rightarrow M}$\\
  \midrule
  $f_{0}(980)\eta$
  & $1.29$ & $2.37$ & $1.29$\\
  $a_{0}(980)\pi$
  & $3.61$ & $2.78$ & $2.78$\\
  $K_{0}^{*}(1430)K$
  & $1.64$ & $2.34$ & $1.64$\\
  $f_{0}(980)\eta+a_{0}(980)\pi$
  & $4.94$ & $5.31$ & $4.94$\\
  $f_{0}(980)\eta+K_{0}^{*}(1430)K$
  & $2.82$ & $4.52$ & $2.82$\\
  $a_{0}(980)\pi+K_{0}^{*}(1430)K$
  & $5.30$ & $5.28$ & $5.28$\\
  all three channels
  & $6.52$ & $7.61$ & $6.52$\\
  \bottomrule
 \end{tabular}
\end{table*}

\subsection{Optimal channel combinations and precision targets}
\label{subsec:optimal-channel-combinations}

The complete three-channel measurement gives the strongest benchmark
separation,
\begin{equation}
 \Delta\chi^2_{H\rightarrow G}=6.52,
 \qquad
 \Delta\chi^2_{H\rightarrow Q}=7.61.
 \label{eq:all-channel-benchmark-result}
\end{equation}
The glueball template is the limiting competitor throughout this benchmark.
These numbers should not be converted directly into discovery significances:
they quantify expected template separation conditional on the adopted
experimental covariance and theory floor.

If only two additional fractions are feasible, the nominal optimum is
\begin{equation}
 S_{\mathrm{best}}^{(2)}
 =
 \left\{
  a_{0}(980)\pi,\,
  K_{0}^{*}(1430)K
 \right\},
 \qquad
 \min_M\Delta\chi^2=5.28.
 \label{eq:benchmark-best-pair}
\end{equation}
For comparison,
\begin{equation}
 S^{(2)}
 =
 \left\{
  f_{0}(980)\eta,\,
  K_{0}^{*}(1430)K
 \right\}
 \quad\Longrightarrow\quad
 \min_M\Delta\chi^2=2.82,
 \label{eq:alternative-pair}
\end{equation}
The $a_0(980)\pi+K_0^*(1430)K$ pair is preferred over this alternative in the
updated symmetric-theory forecast, while $f_0(980)\eta+a_0(980)\pi$ is close
behind at $\min_M\Delta\chi^2=4.94$.  Retaining all three channels is the most
robust design.

At the benchmark $50\%$ theory floor, no tested experimental precision reaches
$\Delta\chi^2=9$: the discrimination is theory limited.  Reducing the
correlated theory uncertainty is therefore at least as important as improving
the experimental precision.

The conclusion is not driven by the benchmark correlation.  Scanning an
equicorrelation over
\begin{equation}
 -0.40\leq\rho\leq0.75
 \label{eq:correlation-scan-range}
\end{equation}
changes the all-channel worst-competitor separation only between
\begin{equation}
 6.51\leq
 \min_M\Delta\chi^2_{H\rightarrow M}
 \leq6.55.
 \label{eq:correlation-scan-result}
\end{equation}
A released covariance remains essential for inference, but an unknown
correlation of plausible magnitude does not erase the forecasted utility of
the three-channel measurement.

\subsection{Flavor cancellation versus a radial node}
\label{subsec:flavor-versus-node-measurement}

Model discrimination and mechanism discrimination are related but distinct
questions.  The conditional one-node emulator separates accepted samples into
a flavor-only region, where the small $K^{*}K$ amplitude is produced mainly
by light--strange flavor interference, and a radial-only region, where it is
produced mainly by an overlap zero.  It uses $6\times10^4$ draws (seed 2371),
$\theta_H\sim U(0^\circ,70^\circ)$, a log-uniform radial scale
$\beta\in[0.45,2.50]~\mathrm{GeV}$, a uniform node momentum in
$[0.35,1.35]~\mathrm{GeV}$, clipped Gaussian SU(3) breaking
$\delta_{SU(3)}\sim N(0,0.20)$ with $|\delta_{SU(3)}|\leq0.60$, and uniform
nuisances for the modeled-width fraction, light-scalar ratio, and scalar-octet
coupling over $[0.50,1]$, $[0.05,0.70]$, and $[0.45,1]$, respectively.
Flavor-only and radial-only samples require suppression ratios below $0.10$
and $0.15$, respectively, while failing the other criterion.  These choices
define the emulator and are not data-derived priors.  The corresponding
median fractions are
\begin{equation}
\begin{split}
 \left(
  \widetilde f_i^{\,\mathrm{flavor}},
  \widetilde f_i^{\,\mathrm{radial}}
 \right)
 ={}&
 \begin{cases}
  (0.1211,\;0.01794),
       & i=f_{0}(980)\eta,\\
  (0.02250,\;0.000592),
       & i=a_{0}(980)\pi,\\
  (0.1364,\;0.3677),
       & i=K_{0}^{*}(1430)K.
 \end{cases}
 \label{eq:mechanism-median-fractions}
\end{split}
\end{equation}
The one-dimensional Jensen--Shannon divergences between the conditional
fraction distributions are
\begin{equation}
 D_{\mathrm{JS}}=
 (0.644,\;0.517,\;0.437)\ {\rm bit}
 \label{eq:mechanism-js-divergences}
\end{equation}
for $f_{0}(980)\eta$, $a_{0}(980)\pi$, and
$K_{0}^{*}(1430)K$, respectively.  Hence
$f_{0}(980)\eta$ is the best individual discriminator inside this emulator.

The momentum ordering explains this result:
\begin{equation}
 p_{f_{0}\eta}=0.879~{\rm GeV},\qquad
 p_{K^{*}K}=0.939~{\rm GeV},\qquad
 p_{a_{0}\pi}=0.970~{\rm GeV},
 \label{eq:near-node-momenta}
\end{equation}
whereas
\begin{equation}
 p_{K_{0}^{*}K}=0.629~{\rm GeV}.
 \label{eq:k0star-momentum}
\end{equation}
A radial zero near the $K^{*}K$ momentum generically suppresses the two
neighboring-momentum scalar channels as well.  Flavor cancellation is instead
specific to the flavor structure of the $K^{*}K$ amplitude and need not
suppress $f_{0}\eta$ or $a_{0}\pi$.  The lower-momentum
$K_{0}^{*}K$ component supplies the complementary check that the expected
hybrid $S+P$ strength survives away from the putative node.

Using a symmetric pooled covariance for the two mechanism medians, the full
three-channel benchmark gives
\begin{equation}
 \Delta\chi^2_{\mathrm{flavor/radial}}=13.19.
 \label{eq:mechanism-all-channel-dchi2}
\end{equation}
For this narrower mechanism question, the best two-channel median-template
combination is $f_{0}(980)\eta+a_{0}(980)\pi$, with
$\Delta\chi^2=10.09$.  This does not contradict the model-comparison ranking:
the two nearby-momentum channels most directly test a common node, while
$K_{0}^{*}K$ is more valuable for separating the complete hybrid pattern
from the glueball and radial-$q\bar q$ templates.

All mechanism sensitivities remain conditional on the one-node emulator,
its parameter priors, and its classification thresholds.  They must not be
interpreted as posterior probabilities for the physical suppression
mechanism.

\begin{figure*}[t]
 \centering
 \includegraphics[width=\textwidth]
 {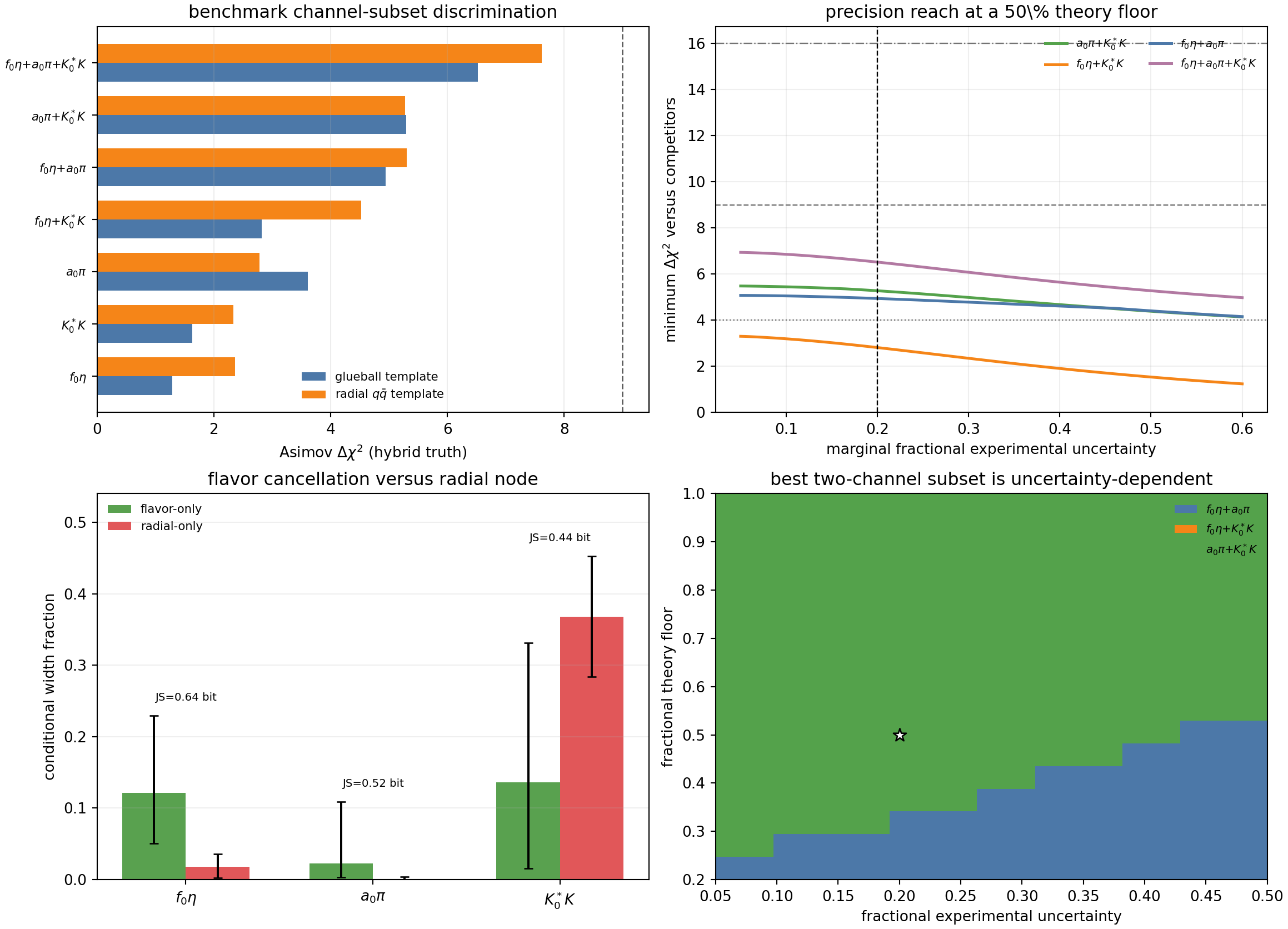}
 \caption{
 Coherent-amplitude measurement design.
 Upper left: benchmark Asimov separations for all channel subsets.
 Upper right: worst-competitor sensitivity as a function of marginal
 experimental precision at a $50\%$ theory floor; the vertical line denotes
 the $20\%$ benchmark.
 Lower left: conditional flavor-only and radial-only width fractions from the
 one-node emulator, with $16$--$84\%$ intervals and the corresponding
 one-dimensional Jensen--Shannon divergences.
 Lower right: the best two-channel subset across the experimental-precision
 and theory-floor scan.  The change of color across the plane demonstrates that
 no pair is universally optimal.  The star denotes the benchmark point.
 No forecast quantity in this figure is included in the current-data score.
 }
 \label{fig:measurement-design}
\end{figure*}

\subsection{Required experimental release}
\label{subsec:required-amplitude-release}

A publication-quality test should be based on a common amplitude analysis,
rather than on independently background-subtracted subchannel yields.  The
minimum useful release consists of the following:

\begin{enumerate}
 \item The complex coefficients for
 $f_{0}(980)\eta$, $a_{0}(980)\pi$,
 $K_{0}^{*}(1430)K$, $K^{*}(892)K$, and the important tensor and
 nonresonant components, quoted as either real and imaginary parts or
 magnitude and phase in an explicit convention.

 \item The complete statistical covariance and a decomposed systematic
 covariance, including correlations between fit fractions, interference
 fractions, pole parameters, efficiency parameters, and shared branching
 normalizations.

 \item Efficiency maps, normalization integrals, or an equivalent
 acceptance-aware likelihood that permits alternative amplitude templates to
 be evaluated without reconstructing the detector model from projected
 histograms.

 \item Alternative scalar line shapes and isobar models, especially for
 $f_{0}(980)$, $a_{0}(980)$, and $K_{0}^{*}(1430)$, together with the
 resulting covariance or likelihood samples.  Variations of nonresonant,
 tensor, and crossed-channel components are required because leakage among
 these waves is a physical model uncertainty.

 \item A common $X(2370)$ pole description across the
 $\pi\pi\eta$ and $K\bar K\pi$ final states.  The joint likelihood should
 share the pole position and appropriate production parameters, but should
 not impose an arbitrary relative phase between physically distinct final
 states.

 \item Preferably, likelihood scans, bootstrap replicas, or posterior samples
 for the complex coefficients.  These are more reusable than a Gaussian
 covariance when the likelihood contains phase ambiguities, boundary effects,
 or multiple amplitude solutions.
\end{enumerate}

The $\pi^{0}\pi^{0}\eta$ sample is particularly valuable because
$f_{0}(980)\eta$ and $a_{0}(980)\pi$ can interfere within the same final
state.  The $K\bar K\pi$ samples provide the complementary
$K^{*}(892)K$ and $K_{0}^{*}(1430)K$ amplitudes.  Combining these samples at
the likelihood level allows their common pole and shared systematic
uncertainties to be propagated without pretending that amplitudes belonging
to different final states interfere.

\subsection{Decision criteria and impact on the composition claim}
\label{subsec:measurement-decision-criteria}

The qualitative mechanism tests are compact.  Suppressed $K^{*}K$ with
$f_0\eta$ and $a_0\pi$ not jointly suppressed favors flavor cancellation
over a common radial node.  Joint suppression of all three makes a radial
node plausible.  A sizable lower-momentum $K_0^*K$ amplitude tests whether
the expected hybrid $S+P$ strength survives away from that node.

These are pattern-level statements and must be tested with the coherent
likelihood, including relative phases and scalar-model uncertainty.

Once the experimental amplitude likelihood becomes available, the final
comparison should replace the synthetic forecast by
\begin{equation}
 {\cal L}_{M}^{\rm total}
 =
 {\cal L}_{M}^{\rm current}
 {\cal L}_{M}^{\rm amplitude},
 \label{eq:future-total-likelihood}
\end{equation}
and repeat the hybrid, pure-glueball, mixed-glueball, and radial-$q\bar q$
comparison with the same nuisance freedom and correlated theory covariance.
The forecasted $\Delta\chi^2$ values must not simply be added to the present
log scores.

Even a decisive separation of $\boldsymbol f_H$ from
$\boldsymbol f_G$ would not distinguish pure from mixed glueball in the
current implementation, because those hypotheses share one decay template.
That remaining ambiguity requires a dedicated mixed-state decay calculation,
the phase-sensitive radiative-production information developed above, or
additional channels whose amplitudes depend differently on the gluonic and
charm components.

The present results are therefore sufficient for the stated scope of this
paper: a flavor-singlet radial hybrid is compatible with current data, but is
not identifiable against a mixed-glueball interpretation, and specific
coherent-amplitude measurements could help resolve much of that ambiguity.  They are
not sufficient for a claim that the $X(2370)$ is predominantly a hybrid.
Such a claim still requires both the experimental likelihood described here
and microscopic hybrid, glueball-mixing, and radial-$q\bar q$ decay
calculations with comparable correlated theory uncertainties.

\section{Discussion, limitations, and interpretation}
\label{sec:discussion}

The preceding sections address three logically distinct questions.  First, can
the present $X(2370)$ observations be accommodated by a flavor-singlet radial
hybrid?  Second, do the current data identify that interpretation relative to
glueball, mixed-glueball, or conventional radial-$q\bar q$ alternatives?
Third, which future measurements could help resolve the remaining ambiguity?  The
answers have different epistemic strengths and should not be combined into a
single composition probability.

Within the signed radial-overlap model, the hybrid interpretation is compatible
with the small visible $K^{*}(892)K$ rate and with the measured total width.
Within the common-nuisance comparison, however, the radial hybrid and
mixed-glueball proxy remain tied.  The proposed coherent quasi-two-body
measurements have substantial forecasted discrimination power between the
hybrid and the present glueball and radial-$q\bar q$ decay templates, but
those amplitudes have not yet been publicly released.  Moreover, because the
pure- and mixed-glueball hypotheses presently share one decay template, the
proposed fractions cannot by themselves distinguish those two interpretations.

\subsection{Hierarchy of the results}
\label{subsec:result-hierarchy}

For clarity, we classify the results into four categories.

\begin{enumerate}
 \item \emph{Direct experimental inputs} are the measured mass, total width,
 inclusive three-pseudoscalar product branching fractions, and visible
 $K^{*}(892)K$ result.  The radiative-production rate used in the likelihood
 is phenomenologically inferred rather than a standalone direct measurement.

 \item \emph{Conditional compatibility statements} test whether a specified
 model template can reproduce those inputs.  The compatibility of the hybrid
 $K^{*}K$ rate and the rescalability of the relative width template belong to
 this category.

 \item \emph{Current identifiability statements} compare all hypotheses with
 the same observable blocks and nuisance construction.  The conclusion that
 the radial hybrid and mixed-glueball proxy are not distinguishable is a result
 of this comparison.

 \item \emph{Prospective forecasts} use synthetic experimental covariance to
 quantify the possible value of future coherent-amplitude measurements.  These forecasts
 are excluded from the current-data score.
\end{enumerate}

This hierarchy prevents two common logical errors.  Compatibility of one model
does not imply exclusion of its competitors, while a forecasted
$\Delta\chi^{2}$ cannot be interpreted as a present experimental preference.

\subsection{The current near-tie is produced by cancellation}
\label{subsec:discussion-cancellation}

Let
\begin{equation}
 \delta_b =
 \log Z_{H,b}-\log Z_{G{\rm -mix},b}
 \label{eq:discussion-block-difference}
\end{equation}
denote the contribution of observable block $b$ to the hybrid-minus-mixed
log score.  At baseline,
\begin{equation}
 \left(
  \delta_{\rm mass},
  \delta_{\Gamma},
  \delta_{\gamma X+K^{*}K},
  \delta_{\rm PPP}
 \right)
 =
 \left(
  -0.1333,\,
  +0.0178,\,
  -0.5594,\,
  +0.7099
 \right).
 \label{eq:discussion-component-vector}
\end{equation}
Their sum is
\begin{equation}
 \Delta\log Z_{H,G{\rm -mix}}
 =
 \sum_b\delta_b
 =
 +0.0350.
 \label{eq:discussion-baseline-dlogz}
\end{equation}
Thus, the small net result does not mean that every input is individually
uninformative.  Radiative production together with the visible $K^{*}K$
constraint favors the mixed-glueball proxy, whereas the inclusive
three-pseudoscalar composition favors the hybrid template.

To quantify this distinction, we define the descriptive cancellation index
\begin{equation}
 {\cal I}_{\rm cancel}
 =
 1-
 \frac{
  \left|\sum_b\delta_b\right|
 }{
  \sum_b|\delta_b|
 }.
 \label{eq:cancellation-index}
\end{equation}
It satisfies $0\leq{\cal I}_{\rm cancel}\leq1$: a value near zero indicates
contributions with a common direction, while a value near unity indicates
strong cancellation.  For the baseline comparison,
\begin{equation}
 {\cal I}_{\rm cancel}=0.9753.
 \label{eq:cancellation-index-result}
\end{equation}
This index is not a likelihood-ratio statistic and has no probabilistic
interpretation.  It simply makes explicit why a new observable can materially
change the model ordering even though the present net score difference is
small.

The corresponding leave-one-block results are
\begin{align}
 \Delta\log Z_{H,G{\rm -mix}}^{(-{\rm mass})}
 &=+0.1683, \nonumber\\
 \Delta\log Z_{H,G{\rm -mix}}^{(-\Gamma)}
 &=+0.0172, \nonumber\\
 \Delta\log Z_{H,G{\rm -mix}}^{(-\gamma X-K^{*}K)}
 &=+0.5945, \nonumber\\
 \Delta\log Z_{H,G{\rm -mix}}^{(-{\rm PPP})}
 &=-0.6749.
 \label{eq:discussion-lobo-results}
\end{align}
Removing either of the two dominant, oppositely directed blocks therefore
changes the apparent preference.  The comparison is controlled by production
and decay composition, not by the mass or total width alone.

The common nuisance scenarios reinforce this conclusion:
\begin{equation}
 \Delta\log Z_{H,G{\rm -mix}}
 =
 \begin{cases}
 -1.289, & \text{tight nuisances},\\
 +0.0350, & \text{baseline nuisances},\\
 +0.415, & \text{wide nuisances}.
 \end{cases}
 \label{eq:discussion-three-scenario-result}
\end{equation}
A continuous shared rescaling of the baseline nuisance widths reverses the
ordering at
\begin{equation}
 \lambda_{\rm cross}=0.9747.
 \label{eq:discussion-common-crossing}
\end{equation}
Only a $2.5\%$ common narrowing is therefore required to exchange the
hybrid and mixed-glueball ordering.  Equal-model-weight normalized score
weights from any one scenario must not be presented as physical composition
probabilities.

\subsection{What hybrid compatibility does and does not mean}
\label{subsec:meaning-of-hybrid-compatibility}

The exact flavor zero inherited from the reference hybrid amplitudes occurs at
\begin{equation}
 \theta_{K^{*}K}^{(0)}=32.31^{\circ},
 \label{eq:discussion-kstar-zero}
\end{equation}
only
\begin{equation}
 \theta_{1}-\theta_{K^{*}K}^{(0)}
 =2.95^{\circ}
 \label{eq:discussion-zero-offset}
\end{equation}
below the flavor-singlet angle
$\theta_{1}=35.26^{\circ}$.  At the singlet point, the signed radial template
gives
\begin{equation}
 f_{K^{*}K}^{H}=1.44\times10^{-4}
 \label{eq:discussion-kstar-fraction}
\end{equation}
and, for the working production branching fraction,
\begin{equation}
 {\cal B}(J/\psi\to\gamma X)
 {\cal B}_{\rm vis}(X\to K^{*}K)
 =
 8.14\times10^{-8}.
 \label{eq:discussion-visible-kstar}
\end{equation}
This is only
\begin{equation}
 \frac{8.14\times10^{-8}}{2.7\times10^{-6}}
 =3.02\times10^{-2}
 \label{eq:discussion-kstar-limit-ratio}
\end{equation}
of the published $90\%$ upper limit.  The null result is therefore
comfortably compatible with the singlet-hybrid template.  It does not,
however, measure the flavor angle or prove that flavor cancellation is the
physical suppression mechanism; a radial node can produce a similar small
rate.

The detailed overlap calculation yields an unscaled modeled sum of
$42.34~{\rm MeV}$; its illustrative MeV column is obtained only after imposing
saturation of the measured $170~{\rm MeV}$ width.  It therefore establishes
that the relative template can be normalized to the observed width scale, not
that $170~{\rm MeV}$ has been predicted.  Scalar assignments, the radial
scale, omitted coupled-channel effects, and the scale-sensitive
$a_{2}(1320)\pi$ channel remain important uncertainties.

The mass supplies a further qualification.  Constituent-gluon and lattice
studies generally place the lowest nonexotic $0^{-+}$ hybrid below the
$X(2370)$ \cite{FarinaSwanson2024HybridDecays}.  Relative to the
$1.75$--$1.90~{\rm GeV}$ lowest-hybrid range, the $X(2370)$ lies
approximately $0.46$--$0.61~{\rm GeV}$ higher.  The proposed assignment
must therefore be radial, higher, or strongly dressed; it cannot be identified
with the lowest state without additional dynamics.  Current-dependent QCD
sum-rule results can overlap the observed mass but do not provide the required
radial wave function or decay amplitudes \cite{TanEtAl2024HybridSumRules}.

\subsection{Radiative production remains a shared bottleneck}
\label{subsec:discussion-radiative-bottleneck}

The working production rate
\begin{equation}
 {\cal B}_{\rm work}(J/\psi\to\gamma X)=3.4\times10^{-3}
 \label{eq:discussion-working-production}
\end{equation}
requires a pseudoscalar transition matrix element
\begin{equation}
 |{\cal M}_{X}(0)|=0.03220~{\rm GeV}^{-1}.
 \label{eq:discussion-required-matrix-element}
\end{equation}
Continuing the $N_f=2$ pseudoscalar-singlet lattice anchor with fixed
dimensionless $V(0)$ gives
\begin{equation}
 {\cal B}_{V{\rm -fixed}}(J/\psi\to\gamma X)
 =1.7746\times10^{-4},
 \qquad
 \kappa_{\rm eff}=4.377
 \label{eq:discussion-fixed-v-result}
\end{equation}
in amplitude.  Holding the dimensionful matrix element fixed instead gives
\begin{equation}
 {\cal B}_{M{\rm -fixed}}(J/\psi\to\gamma X)
 =3.6296\times10^{-4},
 \qquad
 \kappa_{\rm eff}=3.061.
 \label{eq:discussion-fixed-m-result}
\end{equation}
The sampled fixed-$V$ effective factor has the $16$--$84\%$ interval
\begin{equation}
 3.73\leq\kappa_{\rm eff}\leq5.00.
 \label{eq:discussion-effective-factor-interval}
\end{equation}
These factors determine the matrix element required by the working rate; they
do not constitute a microscopic prediction of an anomaly enhancement or a
radial-transition overlap.  A dedicated
$J/\psi\to\gamma H(0^{-+})$ calculation including flavor-singlet
annihilation and anomaly effects remains necessary
\cite{JiangEtAl2023Eta2Radiative,ChenEtAl2023HybridRadiative}.

The mixed-glueball mechanism contains a different unresolved parameter.
Using the lattice-motivated glueball and $\eta_c$ radiative anchors, the
working rate is crossed at
\begin{equation}
 \theta_{G\eta_c}=1.083^{\circ}
 \label{eq:discussion-constructive-mixing}
\end{equation}
for constructive interference, or at
\begin{equation}
 \theta_{G\eta_c}=1.925^{\circ}
 \label{eq:discussion-destructive-mixing}
\end{equation}
after the destructive-interference zero
\cite{GuiEtAl2019PseudoscalarGlueball,ChenEtAl2026GlueballEtaCMixing}.  The preferred conditional grid angle also moves
from $0.9^{\circ}$ to $1.8^{\circ}$ as the relative phase is varied.
Radiative rates therefore do not measure a unique mixing angle unless the sign,
phase convention, and mixing prior are supplied by an independent
nonperturbative calculation.

The two interpretations thus face parallel production uncertainties:
\begin{equation}
 {\cal M}_{H}
 =
 \kappa_{A}R_{\gamma}{\cal M}_{\rm singlet}^{\rm cont},
 \qquad
 {\cal M}_{G{\rm -mix}}
 =
 {\cal M}_{G}\cos\theta
 +
 e^{i\phi}{\cal M}_{\eta_c}\sin\theta.
 \label{eq:discussion-production-degeneracy}
\end{equation}
Present rates constrain only the products or coherent sums appearing in
Eq.~\eqref{eq:discussion-production-degeneracy}.  Neither microscopic
mechanism is currently identified.

\subsection{Unequal microscopic maturity of the decay templates}
\label{subsec:template-maturity}

Giving all hypotheses the same nuisance widths is necessary for a meaningful
screening comparison, but it does not make their underlying predictions
equally microscopic.

\paragraph{Radial hybrid.}
The hybrid template contains channel-dependent signed overlap integrals,
a first-radial node, flavor phases, and a scan over the radial scale.  It is
nevertheless a separable SHO/Jacobi reduction normalized to published
lowest-hybrid amplitudes \cite{FarinaSwanson2024HybridDecays}.  It is not a solution of the
$2.359~{\rm GeV}$ radial constituent-gluon Hamiltonian with a complete decay
operator.

\paragraph{Pure and mixed glueball.}
The glueball fractions are anchored to an extended linear sigma model
\cite{EshraimEtAl2013GlueballDecay}.  The pure- and mixed-glueball hypotheses presently share
those fractions, while differing primarily in their radiative-production
anchors.  This construction is adequate for testing production--decay
identifiability but cannot predict how an $\eta_c$ admixture changes every
light-hadron decay amplitude and relative phase.

\paragraph{Conventional radial $q\bar q$.}
The radial-$q\bar q$ competitor is retained because high radial
$\eta$- or $\eta'$-like assignments have been investigated in
$^{3}P_{0}$-type decay models \cite{YuEtAl2011RadialEta,ChenPing2011RadialEta}.  In the current
common comparison it lies $5.13$--$12.62$ log-score units below the best
model across the wide-to-tight scenarios.  This is a conditional result for
the adopted surrogate template, not a model-independent exclusion of all
conventional radial states.

The scalar structure introduces a common limitation.  Treating
$f_{0}(980)$, $a_{0}(980)$, and $K_{0}^{*}(1430)$ as fixed daughter
states suppresses uncertainties associated with their internal structure,
coupled-channel line shapes, and rescattering.  These effects can alter both
the absolute fractions and their relative phases.  Publication-level model
comparison ultimately requires correlated theory covariance derived from
alternative scalar and decay-operator calculations, not only independent
fractional nuisance floors.

\subsection{What coherent amplitudes can and cannot resolve}
\label{subsec:discussion-future-discrimination}

For the three-fraction benchmark developed in
Sec.~\ref{sec:experimental-discrimination}, hybrid truth gives
\begin{equation}
 \Delta\chi^2(H,G)=6.52,
 \qquad
 \Delta\chi^2(H,Q)=7.61.
 \label{eq:discussion-prospective-separation}
\end{equation}
Scanning the experimental equicorrelation from $-0.40$ to $0.75$ leaves
the worst-competitor proxy separation between $6.51$ and $6.55$.  Thus a
measurement targeting
\begin{equation}
 \left\{
 f_{0}(980)\eta,\,
 a_{0}(980)\pi,\,
 K_{0}^{*}(1430)K
 \right\}
 \label{eq:discussion-target-channel-set}
\end{equation}
can materially constrain the present templates, but the benchmark is theory
limited and is not itself a forecast for a coherent amplitude likelihood.

In contrast,
\begin{equation}
 \Delta\chi^2
 \left(
  G_{\rm pure},
  G{\rm -}\eta_c
 \right)
 =0
 \label{eq:discussion-pure-mixed-zero}
\end{equation}
for these fractions, because the two glueball hypotheses share the same decay
template.  Decay amplitudes alone will resolve the pure--mixed ambiguity only
after a microscopic mixed-state calculation predicts a distinct channel
pattern.  Otherwise, phase-sensitive radiative information or additional
charm-sensitive observables are required.

The possible future outcomes have clear interpretations:

\begin{enumerate}
 \item If the measured complex amplitudes and covariance favor the hybrid
 pattern over both $G$ and $Q$, the result will support the hybrid decay
 mechanism, but will not by itself establish dominant hybrid composition.

 \item If the amplitudes favor the common glueball pattern, the specific radial
 hybrid template constructed here will be disfavored.  A broader hybrid
 interpretation could survive only if an improved microscopic calculation
 produces a substantially different decay covariance.

 \item If none of the three templates describes the coherent likelihood, the
 appropriate conclusion will be failure of the restricted model set, not
 support for whichever template has the smallest residual.

 \item If the decay data favor the common glueball pattern while radiative
 production remains phase-degenerate, pure and mixed glueball will remain
 unresolved until their decay or production amplitudes are independently
 calculated.
\end{enumerate}

\subsection{Claim-readiness audit}
\label{subsec:claim-readiness}

Table~\ref{tab:claim-readiness} summarizes the present status of the analysis.
The labels ``complete,'' ``partial,'' and ``missing'' describe workflow and
input availability; they are not statistical scores.

\begin{table*}[t]
 \centering
 \scriptsize
 \caption{
 Claim-readiness audit.  ``Complete'' means implemented at the stated
 compatibility-paper level, not exact or first-principles.  ``Partial'' denotes
 a controlled but non-microscopic template, while ``missing'' denotes an input
 that cannot be reconstructed from the public data or present calculation.
 }
 \label{tab:claim-readiness}
 \begin{tabular}{p{0.31\textwidth}p{0.10\textwidth}p{0.49\textwidth}}
  \toprule
  Requirement & Status & Interpretation\\
  \midrule
  Public input and covariance audit
   & Complete
   & Measured, inferred, derived, forecast, and unavailable quantities are
     explicitly separated.\\
  Common-nuisance four-model comparison
   & Complete
   & Hybrid, pure glueball, mixed glueball, and radial $q\bar q$ use common
     observable blocks and nuisance scenarios.\\
  Current hybrid compatibility test
   & Complete
   & The visible $K^{*}K$, total width, radiative production, and inclusive
     three-pseudoscalar information are included conditionally.\\
  \midrule
  Microscopic radial-hybrid decay amplitudes
   & Partial
   & Signed, channel-dependent overlaps are implemented, but the radial state
     is a separable surrogate rather than a solved wave function.\\
  Microscopic mixed-glueball decay template
   & Partial
   & Production mixing is phase-aware, but the pure and mixed hypotheses share
     one light-hadron decay template.\\
  Microscopic radial-$q\bar q$ template
   & Partial
   & The conventional competitor is literature-informed but not calculated
     with the same microscopic machinery and covariance.\\
  \midrule
  Measured coherent quasi-two-body likelihood
   & Missing
   & Complex coefficients or fractions with statistical and systematic
     covariance are not publicly available.\\
  Dedicated $J/\psi\to\gamma H(0^{-+})$ calculation
   & Missing
   & The required matrix element is known, but its anomaly and radial factors
     have not been calculated.\\
  Unitary common-pole coupled-channel analysis
   & Missing
   & Independent isobar or Breit--Wigner summaries do not demonstrate a single
     pole across all final states.\\
  \bottomrule
 \end{tabular}
\end{table*}

\begin{figure*}[t]
 \centering
 \includegraphics[width=\textwidth]
 {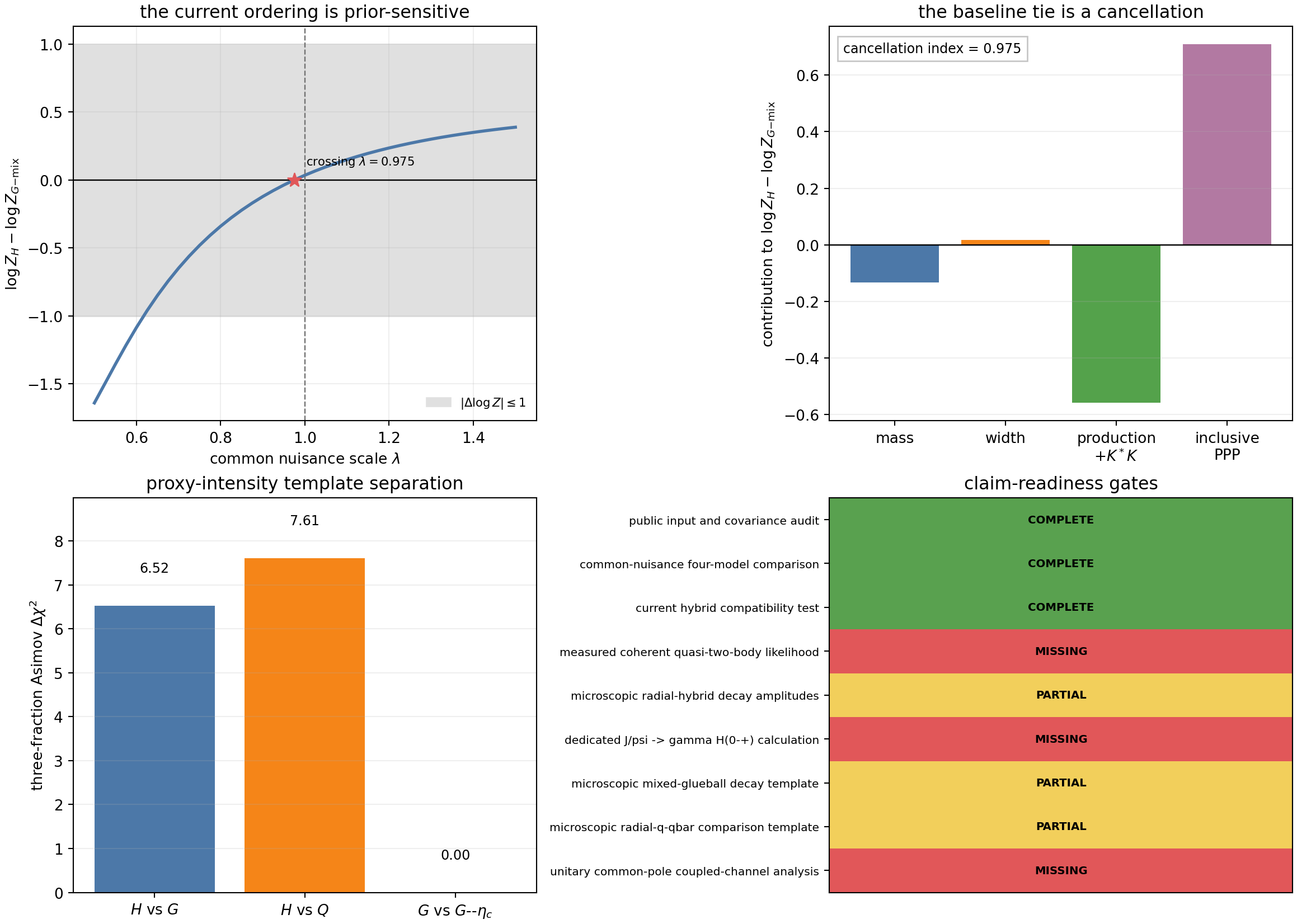}
 \caption{
 Global robustness and claim-readiness synthesis.
 Upper left: continuous common-nuisance dependence of
 $\log Z_H-\log Z_{G{\rm -mix}}$; the star marks the ordering reversal.
 Upper right: baseline observable-block contributions and the descriptive
 cancellation index.
 Lower left: three-proxy-intensity Asimov separation for hybrid versus the shared
 glueball template, hybrid versus radial $q\bar q$, and pure versus mixed
 glueball.  The last separation is identically zero in the present decay
 templates.
 Lower right: status of the nine claim-readiness gates.  These categories
 describe analysis maturity and input availability, not probabilities.
 }
 \label{fig:global-synthesis}
\end{figure*}

\subsection{Scope of the defensible paper claim}
\label{subsec:defensible-paper-claim}

The analysis completed here is sufficient for a phenomenological
compatibility and identifiability paper.  In particular, it supports the
following statement:

\begin{quote}
 A flavor-singlet radial hybrid is compatible with the current $X(2370)$
 data, but remains indistinguishable from a mixed-glueball interpretation
 under the same generic nuisance-width prescription.  Specific amplitude measurements
 could help resolve the hybrid-versus-glueball decay-pattern ambiguity.
\end{quote}

Each part of this statement is quantitatively supported.  Hybrid compatibility
follows from the signed $K^{*}K$ suppression and the rescalability of the
relative width template to the observed scale.
Current indistinguishability follows from the prior-sensitive common comparison
and its $0.9753$ cancellation index.  The proposed experimental resolution
follows from the covariance-aware Asimov forecast.

The present results do \emph{not} support the stronger claims that the
$X(2370)$ is predominantly a hybrid, that the flavor-singlet cancellation
mechanism has been measured, that a glueball interpretation is excluded, or
that a unique glueball--$\eta_c$ mixing angle has been determined.  Those
claims require at minimum:

\begin{enumerate}
 \item a released coherent quasi-two-body likelihood with covariance;
 \item a solved or independently calculated radial-hybrid decay amplitude set;
 \item a dedicated flavor-singlet
 $J/\psi\to\gamma H(0^{-+})$ matrix element;
 \item distinct microscopic decay templates for pure and mixed glueball and
 radial-$q\bar q$ competitors; and
 \item a common-pole coupled-channel analysis across the observed final states.
\end{enumerate}

The correct interpretation of the present work is therefore neither a hybrid
identification nor a negative result.  It demonstrates that the radial-hybrid
possibility survives the strongest presently available constraints, identifies
why the current conditional score cannot distinguish it from a mixed-glueball mechanism,
and converts that ambiguity into a concrete, falsifiable experimental program.

\section{Summary and conclusions}
\label{sec:conclusions}

We tested whether current $X(2370)$ information admits a flavor-singlet
$0^{-+}$ radial-hybrid interpretation when pure glueball, glueball--$\eta_c$
mixing, and radial-$q\bar q$ surrogates are treated with the same likelihood
blocks and generic nuisance-width prescription.  The hybrid construction is an
effective one-coordinate, one-node SHO surrogate anchored to lowest-hybrid
decay widths; it is not a solved two-Jacobi-coordinate radial wave function.
Only the light--strange interference sign within $K^*K$ is inherited from the
published constituent-gluon calculation.  Its small predicted visible
$K^*K$ product, $8.14\times10^{-8}$ at the working production rate, is
therefore a compatibility result rather than a unique hybrid signature.

The overlap weights have an unscaled sum of $42.34~\mathrm{MeV}$ in the
inherited normalization and can be rescaled to the observed
$170^{+44}_{-29}~\mathrm{MeV}$ width.  This is not an absolute width
prediction.  Radiative production is likewise unresolved: the inferred
working rate requires an effective singlet amplitude factor $4.38$ relative
to the fixed-$V(0)$ continuation of the $N_f=2$ lattice anchor, whereas a
small, phase-dependent glueball--$\eta_c$ admixture can reproduce the same
rate.  Neither mechanism is presently calculated well enough to identify the
state's composition.

After positive conditioning of the asymmetric PPP replicas and side-consistent
split-normal residuals, the conditional hybrid-minus-mixed-glueball integrated
scores are $-1.289$, $+0.035$, and $+0.415$ for the tight, baseline, and wide
nuisance scenarios.  The baseline decomposition
$(-0.133,+0.018,-0.559,+0.710)$ for mass, width, production--$K^*K$, and PPP
composition gives a cancellation index of $0.975$.  These values are not
external-prior Bayes factors: several predictive anchors are phenomenological,
the $H$ and $G$ width centers use the observed scale, and the PPP mappings are
fixed central surrogates.  They nevertheless show that the implemented hybrid
and mixed-glueball explanations are not identifiable with current inputs.

The proposed $\{f_0(980)\eta,a_0(980)\pi,K_0^*(1430)K\}$ measurement remains
useful, but the corrected symmetric-theory proxy forecast is more modest than
an exact-template estimate.  With $20\%$ experimental precision, correlation
$0.3$, and a $50\%$ theory floor assigned to both templates, the all-channel
distances are $\Delta\chi^2=6.52$ against the shared glueball template and
$7.61$ against the radial-$q\bar q$ surrogate.  A real test must forward-model
complex amplitudes, line shapes, acceptance, normalization integrals, and
interference; these proxy distances are neither present data nor discovery
significances.  The pure- and mixed-glueball proxies remain indistinguishable
because they share one decay template.

We therefore conclude that a flavor-singlet radial hybrid remains compatible
with current $X(2370)$ data, but is not distinguishable from the implemented
mixed-glueball explanation.  Stronger composition claims require a released
coherent-amplitude likelihood, a full radial-hybrid decay calculation, a
dedicated $J/\psi\to\gamma H(0^{-+})$ matrix element, independently derived
mixed-glueball and radial-$q\bar q$ decay templates, and a common-pole
coupled-channel analysis.

\bibliographystyle{apsrev4-2}
\bibliography{cuboidref-3}

\end{document}